\documentclass[twoside]{article}

\usepackage{graphicx}
\usepackage{subfigure}
\usepackage{amssymb}
\begin{document}

\title{\bf Multiverses and physical cosmology}
\author{G. F. R. Ellis,$^1$ U. Kirchner,$^1$ and W. R. Stoeger$^{1,2}$ \\
\\ \small
$^1$ Department of Mathematics and Applied Mathematics,
University of Cape Town
\\ \small 7700 Rondebosch, South Africa\\
\small
$^2$ Permanent Address: Vatican Observatory Research Group, Steward
Observatory
\\ \small The University of Arizona, Tucson,
 Arizona 85721, USA}
\date{20. June 2003}
\maketitle

\small

\begin{abstract}
The idea of a multiverse -- an ensemble of universes -- has received
increasing attention in cosmology, both as the outcome of the originating
process that generated our own universe, and as an explanation for why our
universe appears to be fine-tuned for life and consciousness. Here we
carefully consider how multiverses should be defined, stressing the
distinction between the collection of all possible universes, and ensembles
of really existing universes that are essential for an anthropic
argument. We show that such realised multiverses are by no means unique. A
proper measure on the space of all really existing universes or universe
domains is needed, so that probabilities can be calculated, and major
problems arise in terms of realised infinities. As an illustration we
examine these issues in the case of the set of 
Friedmann-Lema\^{\i}tre-Robertson-Walker (FLRW) universes. Then we briefly 
summarise scenarios
like chaotic inflation, which suggest how ensembles of universe domains may
be generated, and point out that the regularities which must underlie any
systematic description of truly disjoint multiverses must imply some kind of
common generating mechanism. Finally, we discuss the issue of testability,
which underlies the question of whether multiverse proposals are really
scientific propositions.
\end{abstract}


\section{Introduction}

The idea of a multiverse has been proposed as the only scientifically based
way of avoiding the fine-tuning required to set up the conditions for our
seemingly very unlikely universe to exist. Stephen Weinberg (2000), for
example, uses it to explain the value of the cosmological constant, which he
relates to anthropic issues. Martin Rees (2001) employs it to explain the
whole set of anthropic coincidences (Barrow and Tipler 1986), that is, to
explain why our universe is a congenial home for life. These and similar
proposals have been triggered by the dawning awareness among many
researchers that there may be many other existing universes besides ours.
This possibility has received strong stimulation from proposals like Andrei
Linde's (1983, 1990) chaotic inflationary scenario, in which the origin of
our own observable universe region naturally involves the origin of many
other similar expanding universe regions.

There is however a vagueness about the proposed nature of multiverses. They
might occur in various ways, discussed by Weinberg (2000) and Tegmark
(2003). They might originate naturally in different times and places through
meta-cosmic processes like chaotic inflation, or in accord with Lee Smolin's
(1999) cosmic Darwinian vision. In the latter case, an ensemble of expanding
universe regions grow from each other following gravitational collapse and
re-expansion, where natural selection of universes through optimisation of
black hole production leads to bio-friendly universe regions. This is an
intriguing idea, but with many uncertain steps -- in particular no proof has
been given of the last step, that the physics that maximises black hole
production also favours life. They might be associated with the
multi-universe Everett-Wheeler-type interpretation of quantum mechanics. Or
perhaps multiverses can be truly disjoint collections of universes (see
Sciama 1993, Rees 2001, Tegmark 1998, 2003).

Some refer to the separate expanding universe regions in chaotic inflation
as `universes', even though they have a common causal origin and are all
part of the same single spacetime. In our view (as `uni' means `one') 
\textit{the Universe} is by definition the one unique connected 
\footnote{\textquotedblleft Connected\textquotedblright\ implies 
\textquotedblleft
Locally causally connected\textquotedblright , that is all universe domains
are connected by $C^{0}$ timelike lines which allow any number of reversals
in their direction of time, as in Feynman's approach to electrodynamics.
Thus it is a union of regions that are causally connected to each other, and
transcends particle and event horizons; for examples all points in de Sitter
space time are connected to each other by such lines.} existing spacetime of
which our observed expanding cosmological domain is a part. We will refer to
situations such as in chaotic inflation as a \textit{Multi-Domain Universe},
as opposed to a completely causally disconnected \textit{Multiverse}.
Throughout this paper, when our discussion pertains equally well to disjoint
collections of universes (multiverses in the strict sense) and to the
different domains of a Multi-Domain Universe, we shall for simplicity simply
use the word \textquotedblleft \textit{ensemble}\textquotedblright . When an
ensemble of universes are all sub-regions of a larger connected spacetime -
the \textquotedblleft universe as a whole\textquotedblright - we have the
multi-domain situation, which should be described as such. Then we could
reserve \textquotedblleft multiverse\textquotedblright\ for the collection
of genuinely disconnected \textquotedblleft universes\textquotedblright\ --
those which are not locally causally related.

So far, none of these ideas have been developed to the point of actually
describing such ensembles of universes in detail, nor has it been
demonstrated that a generic well-defined ensemble will admit life. Some
writers tend to imply that there is only one possible multiverse
(characterised by ``all that can exist does exist''). This vague
prescription actually allows a vast variety of different realisations with
differing properties, leading to major problems in the definition of the
ensembles and in averaging, due to the lack of a well-defined measure and
the infinite character of the ensemble itself. Furthermore it is not at all
clear that we shall ever be able to accurately delineate the class of all
possible universes.

The aim of this paper is to help clarify what is involved in a full
description of an ensemble of universes. Our first contribution is
clarifying what is required in order to describe the space of possible
universes, where much hinges on what we regard as `possible'. However that
is only part of what is needed. It is crucial to recognise that anthropic
arguments for \textit{existence} based on ensembles of universes with
specific properties require an \textit{actually existing ensemble} with all
the required properties. For purposes of providing an explanation of
existence, simply having a conceptually possible ensemble is not adequate -
one needs a link to objects or things that actually exist, or to mechanisms
that make them exist.

The second contribution of this paper is to show how an actually existing
ensemble may be described in terms of a space of possible universes, by
defining a distribution function (discrete or continuous) on the space of
possible universes. This characterizes which of the theoretically possible
universes have been actualised in the ensemble - it identifies those that
have actually come into existence. This leads us to our third point: the
problems arising when it is claimed that there is an actually existing
ensemble containing an infinite number of universes or of expanding universe
regimes. Actually existing infinities are very problematic.

There are fundamental issues that arise in considering ensembles of actually
existing universes: what would explain the existence of an ensemble, and its
specific properties? Why should there be this particular ensemble, rather
than some other one? Why should there be any regularity at all in its
properties? The fourth point we make is that if all the universes in an
ensemble show regularities of structure, then that implies some common
generating mechanism.\ Some such structuring is necessary if we are to be
able to describe a multiverse with specified properties - a coherent
description is only possible through the existence of such regularities.
Hence a multiverse consisting of completely causally disconnected universes
is a problematic concept.

The issue of testability is a further important consideration: Is there any
conceivable direct or indirect way of testing for existence of an ensemble
to which our universe belongs? Our fifth point is that there is no way we
can test any mechanism proposed to impose such regularities:\ they will of
necessity always remain speculative. The sixth point is to argue that
existence of multiverses or ensembles is in principle untestable by any
direct observations, and the same applies to any hypothesized properties we
may suppose for them. However certain observations would be able to disprove
existence of some multi-domain ensembles. It is only in that sense that the
idea is a testable proposition.

It is clear that in dealing with multiverses one inevitably runs up against
philosophical and metaphysical issues, for example concerning the ability to
make scientific conclusions in the absence of observational evidence, and in
pursuing the issue of realised infinities. A companion more philosophically
oriented paper will pursue those issues.

\section{Describing Ensembles: Possibility}

To characterise an ensemble of existing universes, we first
need to develop adequate methods for describing the class of all possible
universes. This requires us to specify, at least in principle, all the
ways in which universes can be different from one another, in terms of
their physics, chemistry, biology, etc. 

\subsection{The Set of Possible Universes}

The basis for describing ensembles or multiverses is contained in the
structure and the dynamics of a space $\mathcal{M}$ of all possible
universes $m$, each of which can be described in terms of a set of states $s$
in a state space $\mathcal{S}$. Each universe in $\mathcal{M}$ will be
characterised by a set $\mathcal{P}$ of distinguishing parameters $p$, which
are coordinates on $\mathcal{S}$. Some will be logical parameters, some will
be numerical constants, and some will be functions or tensor fields defined
in local coordinate neighbourhoods for $s$. Each universe $m$ will evolve
from its initial state to some final state according to the dynamics
operative, with some or all of its parameters varying as it does so. The
course of this evolution of states will be represented by a path in the
state space $\mathcal{S}$, depending on the parametrisation of $\mathcal{S}$
. Thus, each such path (in degenerate cases a point) is a representation of
one of the universes $m$ in $\mathcal{M}$. The coordinates in $\mathcal{S}$
will be directly related to the parameters specifying members of $\mathcal{M}
$. The parameter space $\mathcal{P}$ has dimension $N$ which is the
dimension of the space of models $\mathcal{M}$; the space of states $
\mathcal{S}$ has $N+1$ dimensions, the extra dimension indicating the change
of each model's states with time, characterised by an extra parameter, e.g.,
the Hubble parameter $H$ which does not distinguish between models but
rather determines what is the state of dynamical evolution of each model.
Note that $N$ may be infinite, and indeed will be so unless we consider only
geometrically highly restricted sets of universes.

It is possible that with some parameter choices the same physical universe $
m $ will be multiply represented by this description; thus a significant
issue is the equivalence problem -- identifying which different
representations might in fact represent the same universe model. In
self-similar cases we get a single point in $\mathcal{S}$ described in terms
of the chosen parameters $\mathcal{P}$: the state remains unchanged in terms
of the chosen variables. But we can always get such variables for any
evolution, as they are just comoving variables, not necessarily indicating
anything interesting is happening dynamically. The interesting issue is if
this invariance is true in physically defined variables, e.g., expansion
normalised variables, then physical self-similarity is occurring.

The very description of this space $\mathcal{M}$ of possibilities is based
on an assumed set of laws of behaviour, either laws of physics or meta-laws
that determine the laws of physics, which all universes $m$ have in common;
without this, we have no basis for setting up its description. The detailed
characterisation of this space, and its relationship to $\mathcal{S}$, will
depend on the matter description used and its behaviour. The overall
characterisation of $\mathcal{M}$ therefore must incorporate a description
both of the geometry of the allowed universes and of the physics of matter.
Thus the set of parameters $\mathcal{P}$ will include both geometric and
physical parameters.

The space $\mathcal{M}$ has a number of important subsets, for example:

\begin{enumerate}
\item  $\mathcal{M}_{\mathrm{FLRW}}$ -- the subset of all possible exactly
Friedmann-Lema\^{\i}tre-Robertson-Walker (FLRW) universes, described by the
state space $\mathcal{S}_{\mathrm{FLRW}}$ (in the case of dust plus
non-interacting radiation a careful description of this phase space has been
given by Ehlers and Rindler1989).

\item  $\mathcal{M}_{\mathrm{almost-FLRW}}$ -- the subset of all perturbed
FLRW model universes. These need to be characterised in a gauge-invariant
way (see e.g. Ellis and Bruni 1989) so that we can clearly identify those
universes that are almost-FLRW\ and those that are not.

\item  $\mathcal{M}_{\mathrm{anthropic}}$ -- the subset of all possible
universes in which life emerges at some stage in their evolution. This
subset intersects $\mathcal{M}_{\mathrm{almost-FLRW}}$, and may even be a
subset of $\mathcal{M}_{\mathrm{almost-FLRW}},$ but does not intersect $
\mathcal{M}_{\mathrm{FLRW}}$ (realistic models of a life-bearing universe
like ours cannot be exactly FLRW, for then there is no structure).

\item  $\mathcal{M}_{\mathrm{Observational}}$ -- the subset of models
compatible with current astronomical observations. Precisely because we need
observers to make observations, this is a subset of $\mathcal{M}_{\mathrm{\
anthropic}}$.
\end{enumerate}

If $\mathcal{M}$ truly represents all possibilities, one must have a
description that is wide enough to encompass \textit{all} possibilities. It
is here that major issues arise: how do we decide what all the possibilities
are? What are the limits of possibility? What classifications of possibility
are to be included? \textquotedblleft All that can happen happens
\textquotedblright\ must imply all possibilities, as characterised by our
description in terms of families of parameters:\ all allowed values must
occur, and they must occur in all possible combinations. The full space $
\mathcal{M}$ must be large enough to represent all of these possibilities,
along with many others we cannot even conceive of, but which can
nevertheless in principle also be described by such parameters. An
interesting related point has been pointed to us by Jean-Phillipe Uzan:\ it
may be that the larger the possibility space considered, the more fine-tuned
the actual universe appears to be - for with each extra possibility that is
included in the possibility space, unless it can be shown to relate to
already existing parameters, the actual universe and its close neighbours
will live in a smaller fraction of the possibility space. For example if we
assume General Relativity then there is only the parameter $G$ to measure;
but if we consider  scalar-tensor theories, then we have to explain why we
are so close to General Relativity now. Hence there is a tension between
including all possibilities in what we consider, and giving an explanation
for fine tuning.

From these considerations we have the first key issue: \newline

\noindent \textbf{Issue 1:} What determines $\mathcal{M}$? Where does this
structure come from? What is the meta-cause that delimits this set of
possibilities? Why is there a uniform structure across all universes $m$ in $
\mathcal{M}$?

The meta-question is whether any of these questions can be answered
scientifically. We return to that at the end.

\subsection{Adequately Specifying Possible Anthropic Universes}

When defining any ensemble of universes, possible or realised, we must
specify all the parameters which differentiate members of the ensemble from
one another at any time in their evolution. The values of these parameters
may not be known or determinable initially in many cases -- some of them may
only be set by transitions that occur via processes like symmetry breaking
within given members of the ensemble. In particular, some of the parameters
whose values are important for the origination and support of life may only
be fixed later in the evolution of universes in the multiverse.

We can separate our set of parameters $\mathcal{P}$ for the space of all
possible universes $\mathcal{M}$ into different categories, beginning with
the most basic or fundamental, and progressing to more contingent and more
complex categories. Ideally they should all be independent of one another,
but we will not be able to establish that independence for each parameter,
except for the most fundamental cosmological ones. In order to categorise
our parameters, we can doubly index each parameter $p$ in $\mathcal{P}$ as $
p_j(i)$ such that those for $j=1-2$ describe basic physics, for $j=3-5$
describe the cosmology (given that basic physics), and $j=6-7$ pertain
specifically to emergence and life (we must include the latter if we
seriously intend to address anthropic issues). Our characterisation is as
follows:

\begin{enumerate}
\item  $p_1(i)$ are the basic physics parameters within each universe,
excluding gravity - parameters characterising the basic non-gravitational
laws of physics in action, related constants such as the fine-structure
constant $\alpha $, and including parameters describing basic particle
properties (masses, charges, spins, etc.) These should be logical parameters
or dimensionless parameters, otherwise one may be describing the same
physics in other units.

\item  $p_2(i)$ are basic parameters describing the nature of the
cosmological dynamics, e. g., $p_2(1)=1$ indicates Einstein gravity
dominates, $p_2(1)=2$ indicates Brans-Dicke theory dominates, $p_2(1)=3$
indicates Electro-magnetism dominates, etc. Associated with each choice are
the relevant parameter values, e.g., $p_2(2)=G$, $p_2(3)=\Lambda $, and in
the Brans-Dicke case $p_2(4)=\omega $. If gravity can be derived from more
fundamental physics in some unified fundamental theory, these will be
related to $p_1(i);$ for example the cosmological constant may be determined
from quantum field theory and basic matter parameters.

\item  $p_3(i)$ are cosmological parameters characterising the nature of the
matter content of a universe. These parameters encode whether radiation,
baryons, dark matter, neutrinos, scalar fields, etc. occur, in each case
specifying the relevant equations of state and auxiliary functions needed to
determine the physical behaviour of matter (e.g. barotropic equations of
state and the potential function for scalar fields). These are
characterisations of physical possibilities for the macro-states of matter
arising out of fundamental physics, so the possibilities here will be
related to the parameters in $p_1(i)$. Realistic representations will
include all the above, but simplified ensembles considered for exploratory
purposes may exclude some or many of them.

\item  $p_4(i)$ are physical parameters determining the relative amounts of
each kind of matter present in the specific cosmological solutions
envisaged, for example the density parameters $\Omega _{i}$of
various components at some specific stage of its evolution (which then for
example determine the matter to anti-matter ratio and the entropy to baryon
ratio). The matter components present will be those characterised by $p_3(i).
$

\item  $p_5(i)$ are geometrical parameters characterising the spacetime
geometry of the cosmological solutions envisaged- for example the scale
factor $a(t),$ Hubble parameter $H(t)$, and spatial curvature parameter $k$
in FLRW\ models. These will be related to $p_4(i)\,$by the gravitational
equations set in $p_2(i),$ for example the Einstein Field Equations.

\item  $p_6(i)$ are parameters related to the functional emergence of
complexity in the hierarchy of structure, for example allowing the existence
of chemically complex molecules. Thus $p_6(1)$ might be the number of
different types of atoms allowed (as characterised in the periodic table), $
p_6(2)$ the number of different states of matter possible (solid, liquid,
gas, plasma for example), and $p_6(3)$ the number of different types of
molecular bonding. These are emergent properties arising out of the
fundamental physics in operation, and so are related to the parameters set
in $p_1(i)$.

\item  $p_7(i)$ are biologically relevant parameters related specifically to
the functional emergence of life and of self-consciousness, for example $
p_7(1)$ might characterise the possibility of supra-molecular chemistry and $
p_7(2)$ that of living cells. This builds on the complexity allowed by $
p_6(i)$ and relates again to the parameter set $p_1(i)$

It is important to note that these parameters will describe the set of
possibilities we are able to characterise on the basis of our accumulated
scientific experience. The limits of our understanding are relevant here, in
the relation between what we conceive of as this space of possibilities, and
what it really is. There may be universes which we believe are possible on
the basis of what we know of physics, that may in fact not be possible.
There may also be universes which we conceive of as being impossible for one
reason or another, that turn out to be possible. And it is very likely that
we simply may not be able to imagine or envisage all the possibilities.
However this is by no means a statement that ``all that can occur'' is
arbitrary. On the contrary, specifying the set of possible parameters
determines a uniform high-level structure that is obeyed by all universes in 
$\mathcal{M}$.

We see, then, that a possibility space $\mathcal{M\,}$ is the set of
universes (one-parameter sets of states $\mathcal{S}$) obeying the dynamics
characterised by a parameter space $\mathcal{P}$, which may be considered to
be the union of all allowed parameters $p_j(i)$ for all $i,j\,\,$as briefly
discussed above: 
\[
\mathcal{M}=\{\mathcal{S},\mathcal{P}\},~~\mathcal{P}=\cup
_{i,j}\,p_j(i).
\]
Because the parameters $\mathcal{P}$ determine the dynamics, the set of
paths in $\mathcal{S}$ characterising individual universes $m$ are
determined by this prescription. In some particular envisaged ensemble, some
of these parameters (`class parameters') may be fixed across the ensemble,
thus defining a class of universes considered, while others (`member
parameters') will vary across the ensemble, defining the individual members
of that class. Thus 
\[
\mathcal{P=P}_{class}\mathcal{\cup ~P}_{member}.
\]
As we consider more generic ensembles, class parameters will be allowed to
vary and so will become member parameters. In an ensemble in which all that
is possible happens, all parameters will be member parameters; however that
is so hard to handle that we usually analyse sub-spaces 
characterised by particular class parameters.
\end{enumerate}

\subsection{Describing the Geometry of Possible Universes}

Cosmological models are characterised by a preferred timelike vector field $
u:~u^{a}u_{a}=-1,$ usually the fluid flow vector (Ellis 1971a), but sometimes
chosen for other reasons, e.g. to fit local symmetries. To describe a
cosmological spacetime locally we must give a description of its (generally
inhomogeneous and anisotropic) geometry via suitable parameters $p_{5}(i)$.
This description may be usefully given in terms of a tetrad basis as follows
(see Ellis and van Elst 1999, Wainwright and Ellis 1996, Uggla, \textit{et
al.} 2003):

\textit{Feature 1}:\ a set of local coordinates $\mathcal{X}$ $=\{x^i\}$
must be chosen in each chart of a global atlas. This will in particular have
a time coordinate $t\,$ which will be used to characterise evolution of the
universe; this should be chosen in as uniform as possible a way across all
the universes considered, for example it may be based on surfaces of
constant Hubble parameter $H$ for the preferred vector field $u$.

\textit{Feature 2}: in each chart, to determine the geometry we must be
given the components $\mathcal{E}=[e_{\;a}^i(x^j)]$ of an orthonormal tetrad
with the fluid flow vector chosen as the timelike tetrad vector ($a,b,c..$
are tetrad indices; four of these components can be set to zero by suitable
choice of coordinates). Together the coordinates and the tetrad form the
reference frame 
\begin{equation}
\mathcal{P}_{\mathrm{frame}}\equiv \{\mathcal{X},\mathcal{E\}}.
\end{equation}
The metric tensor is then 
\[
ds^2=g_{ij}\,(x^k)dx^idx^j=\eta
_{ab}\,e_{\;i}^a(x^k)\,e_{\;j}^b(x^l)dx^idx^j 
\]
where $\eta _{ab}$ is the Minkowski metric:\ 
\[
\eta _{ab}=e_a.e_b=diag(-1,+1,+1,+1) 
\]
(because the tetrad is orthonormal) and $e_{\;j}^b(x^l)$ are the inverse of $
e_{\;a}^i(x^j):$ 
\[
e_{\;a}^i(x^j)e_i^{\;b}(x^j)=\delta _a^b. 
\]
Thus the metric is given by 
\begin{equation}
ds^2=-\left( e_{\;i}^0dx^i\right) ^2+\left( e_{\;i}^1dx^i\right) ^2+\left(
e_{\;i}^2dx^i\right) ^2+\left( e_{\;i}^3dx^i\right) ^2  \label{metric2}
\end{equation}

The basic geometric quantities used to determine the spacetime geometry are
the rotation coefficients $\Gamma _{\;bc}^a$ of this tetrad, defined by 
\[
\Gamma _{\;bc}^a=e_j^{\;a}e_{\;c;k}^je_{\;b}^k. 
\]
They may conveniently be given in terms of geometric quantities 
\begin{equation}
\mathcal{P}_{\mathrm{geometry}}\equiv \{\dot{u}_\alpha ,\theta ,\sigma
_{\alpha \beta },\omega _{\alpha \beta },\Omega _\gamma ,a^\alpha ,n_{\alpha
\beta }\}.  \label{geom}
\end{equation}
characterised as follows:

\begin{eqnarray*}
\Gamma _{\alpha 00} &=&\dot{u}_{\alpha }, \\
\Gamma _{\alpha 0\beta } &=&\frac{1}{3}\theta +\sigma _{\alpha \beta
}-\omega _{\alpha \beta }, \\
\Gamma _{\alpha \beta 0} &=&\epsilon _{\alpha \beta \gamma }\Omega ^{\gamma
}, \\
\Gamma _{\alpha \beta \gamma } &=&a_{[\alpha }\delta _{\beta ]\gamma
}+\epsilon _{\gamma \delta \lbrack \alpha }n_{\;\beta ]}^{\delta }+\frac{1}{
2 }\epsilon _{\alpha \beta \delta }n_{\;\delta }^{\gamma },
\end{eqnarray*}
where $\dot{u}_{\alpha }$ is the acceleration of the fluid flow congruence, $
\theta $ is its expansion, $\sigma _{\alpha \beta }\,=\sigma _{(\alpha \beta
)}$ is its shear $(\sigma _{\;b}^{b}=0),$ and $\omega _{\alpha \beta
}=\omega _{\lbrack \alpha \beta ]}$ its vorticity, while $n_{\alpha \beta
}=n_{(\alpha \beta )}$ and $a_{\alpha }$ determine the spatial rotation
coefficients (see Wainwright and Ellis 1996, Ellis and van Elst 1999). Greek
indices (with range $1-3)$ indicate that all these quantities are orthogonal
to $u^{a}.$ They are spacetime fields, although in particular high-symmetry
cases they may be independent of many or of all the coordinates. The Jacobi
identities, Bianchi identities, and Einstein field equations can all be
written out in terms of these quantities, as can the components $E_{\alpha
\beta },$ $H_{\alpha \beta }$ of the Weyl tensor (see Ellis and van Elst
1999). Except in the special cases of isotropic spacetimes and locally
rotationally symmetric spacetimes (see Ellis 1967, van Elst and Ellis 1996),
the basis tetrad can be chosen in an invariant way so that three of these
quantities vanish and all the rest are scalar invariants.

Thus the geometry is determined by the 36 spacetime functions in the
combined set ($\mathcal{E},\mathcal{P}_{\mathrm{geometry}}$) with some
chosen specification of coordinates $\mathcal{X}$, with the metric then
determined by (\ref{metric2}). For detailed dynamical studies it is often
useful to rescale the variables in terms of the expansion (see Wainwright
and Ellis 1995, Uggla et al 2003 for details). Note that the same universe
may occur several times over in this space; the \textit{equivalence problem}
is determining when such multiple representations occur. We do not recommend
going to a quotient space where each universe occurs only once, as for
example in the dynamical studies of Fischer and Marsden (1979), for the cost
of doing so is to destroy the manifold structure of the space of spacetimes.
It is far better to allow multiple representations of the same universe (for
example several representations of the same Bianchi I universe occur in the
Kasner ring in the space of Bianchi models, see Wainwright and Ellis 1996)
both to keep the manifold structure intact and because then the dynamical
structure becomes clearer.

\textit{Feature 3}:\ To determine the global structure, we need a set of
composition functions relating different charts in the atlas where they
overlap, thus determining the global topology of the universe.

Together these are the parameters $p_5(i)$ needed to distinguish model
states. A\ particular model will be represented as a path through those
states. The nature of that evolution will be determined by the matter
present.

\subsection{Describing the Physics of Possible Universes}

\textit{Feature 4}: To determine the matter stress-energy tensor we must
specify the quantities 
\begin{equation}
\mathcal{P}_{\mathrm{matter}}\equiv \{\mu ,q_\alpha ,p,\pi _{ab},\Phi _A\}
\label{mat}
\end{equation}
for all matter components present, where $\mu $ is the energy density, $
q_\alpha $ is the momentum flux density, $p$ is the pressure, $\pi _{ab}=\pi
_{(ab)}$ the anisotropic pressure $(\pi _{\;b}^b=0),$ and $\Phi _A$ ($
A=1..A_{\max })$ is some set of internal variables sufficient to make the
matter dynamics deterministic when suitable equations of state are added
(for example these might include the temperature, the entropy, the velocity $
v^i$ of matter relative to the reference frame, some scalar fields and their
time derivatives, or a particle distribution function)$.$ These are
parameters $p_4(i)\,$ for each kind of matter characterised by $p_3(i).$
Some of these dynamical quantities may vanish (for example, in the case of a
`perfect fluid', $q_\alpha =0,$ $\pi _{ab}=0)\,$and some of those that
do not vanish will be related to others by the equations of state (for
example, in the case of a barotropic fluid, $p=p(\mu ))$ and dynamic
equations (for example the Klein Gordon equation for a scalar field). These
equations of state can be used to reduce the number of variables in $
\mathcal{P}_{\mathrm{matter}}$; when they are not used in this way, they
must be explicitly stated in a separate parameter space $\mathcal{\ P}_{
\mathrm{eos}}$ in $p_3(i).$ In broad terms 
\begin{eqnarray}
\mathcal{P}_{\mathrm{eos}} &\equiv &\{q_\alpha =q_\alpha (\mu ,\Phi
_A),\;~~p=p(\mu ,\Phi _A),\;  \nonumber \\
&{}&\;\pi _{ab}=\pi _{ab}(\mu ,\Phi _A),\;\dot{\Phi}_A=\dot{\Phi}_A(\Phi
_A)\}.
\end{eqnarray}

Given this information the equations become determinate and we can obtain
the dynamical evolution of the models in the state space; see for example
Wainwright and Ellis (1996), Hewitt et al (2002), Horwood et al (2002) for
the case of Bianchi models (characterised by all the variables defined above
depending on the time only) and Uggla et al (2003), Lim et al (2003) for the
generic case.

\textit{Feature 5}: However more general features may vary: the
gravitational constant, the cosmological constant, and so on; and even the
dimensions of spacetime or the kinds of forces in operation. These are the
parameters $\mathcal{P}_{\mathrm{physics}}$ comprising $p_1(i)$ and $p_2(i).$
What complicates this issue is that some or many of these features may be
emergent properties, resulting for example from broken symmetries occurring
as the universe evolves. Thus they may come into being rather than being
given as initial conditions that then hold for all time.

Initially one might think that considering all possible physics simply
involves choices of coupling constants and perhaps letting some fundamental
constant vary. But the issue is more fundamental than that. Taking seriously
the concept of including \textit{all} possibilities in the ensembles, the
space of physical parameters $\mathcal{P}_{\mathrm{physics}}$ used to
describe $\mathcal{M}$, the parameters $p_{2}(i)$ might for example include
a parameter $p_{\mathrm{grav}}(i)$ such that: for $i=1$ there is no gravity;
for $i=2$ there is Newtonian gravity; for $i=3$ general relativity is the
correct theory at all energies -- there is no quantum gravity regime; for $
i=4$ loop quantum gravity is the correct quantum gravity theory; for $i=5$ a
particular version of superstring theory or M-theory is the correct theory.

Choices such as these will arise for all the laws and parameters of physics.
In some universes there will be a fundamental unification of physics
expressible in a basic ``theory of everything'', in others this will not be
so. Some universes will be realised as branes in a higher dimensional
spacetime, others will not.

\subsection{The Anthropic subset}

We are interested in the subset of universes that allow intelligent life to
exist. That means we need a function on the set of possible universes that
describes the probability that life may evolve. An adaptation of the Drake
equation (Drake and Shostak) gives for the probability of intelligent life
in any particular universe $m$ in an ensemble, 
\begin{equation}
P_{life}(m)=F*\Pi  \label{life1}
\end{equation}
where the existence of a habitat for life is expressed by the product 
\begin{equation}
\Pi =P_{gal}*R*f_S*f_p*n_e  \label{life4}
\end{equation}
and the coming into existence of life, given such a habitat, is expressed by
the product 
\begin{equation}
F=f_l*f_i.  \label{life5}
\end{equation}
Here $P_{gal}\,$is the probability of galaxies forming in the universe
considered, $R$ is the average rate of star formation in galaxies, $f_S$ is
the fraction of these stars that can provide a suitable environment for life
(they are `Sun-like'), $f_p$ is the fraction of stars that are surrounded by
planetary systems, $n_e$ is the mean number of planets in each such system
that are suitable habitats for life (they are `Earth-like'), $f_l$ is the
fraction of such planets on which life actually originates, and $f_i$
represents the fraction of those planets on which there is life where
self-conscious beings develop. The anthropic subset of a possibility space
is that set of universes for which $P_{life}(m)>0.$

The probabilities \{$P_{gal},R,f_S,f_p,n_e,f_l,f_i$) are functions of the
physical and cosmological parameters characterised above, so there will be
many different representations of this parameter set depending on the degree
to which we try to represent such interrelations. The astrophysical issues
expressed in the product $\Pi $ are the easier ones to investigate. We can
in principle make a cut between those consistent with the eventual emergence
of life and those incompatible with it by considering each of the factors in 
$\Pi \,\,$in turn, taking into account their dependence on the parameters $
p_1(i)$ to $p_5(i),$ and only considering the next factor if all the
previous ones are non-zero (an approach that fits in naturally with Bayesian
statistics and the successive allocation of relevant priors). $\,$In this
way we can assign \textquotedblleft bio-friendly intervals\textquotedblright
\ to the possibility space $\mathcal{M}$. If $\ \Pi \,$ is non-zero we can
move on to considering similarly whether $F$ is non-zero, based on the
parameters $p_6(i)$ to $p_7(i)$ determining if true complexity is possible,
which in turn depend on the physics parameters $p_1(i)$ in a crucial way
that is not fully understood. It will be impossible at any stage to
characterise that set of the multiverse in which \textit{all} the conditions 
\textit{necessary} for the emergence of self-conscious life and its
maintenance have been met, for we do not know what those conditions are (for
example, we do not know if there are forms of life possible that are not
based on carbon and organic chemistry). Nevertheless it is clear that life
demands unique combinations of many different parameter values that must be
realised simultaneously. When we look at these combinations, they will span
a very small subset of the whole parameter space (Davies 2003, Tegmark 2003).

If we wish to deal with specifically human life, we need to make the space $
\mathcal{M}$ large enough to deal with all relevant parameters for this
case, where free will arises. This raises substantial extra complications,
discussed in the companion (more philosophical) paper.

\subsection{Parameter space revisited}

It is now clear that some of the parameters discussed above are dependent on
other ones, so that while we can write down a more or less complete set at
varying levels of detail they will in general not be an independent set.
There is a considerable challenge here: to find an independent set. \textit{
Inter alia} this involves solving both the initial value problem for general
relativity and the way that galactic and planetary formation depend on
fundamental physics constants (which for example determine radiation
transfer properties in stars and in proto-planetary gas clouds), as well as
relations there may be between the fundamental constants and the way the
emergent complexity of life depends on them. We are a long way from
understanding all these issues. This means we can provide necessary sets of
parameter values but cannot guarantee completeness or independence.

\section{The Set of Realised Universes}

We have now characterised the set of possible universes. But in any given
ensemble, they may not all be realised, and some may be realised many times.
The purpose of this section is to set up a formalism making clear which of
the \textit{possible} universes (characterised above) 
occur in a specific \textit{realised} ensemble.

\subsection{A distribution function describing an ensemble of realised
universes}

In order to select from $\mathcal{M}$ a set of realised universes we need to
define on $\mathcal{M}$ a distribution function $f(m)$ specifying how many
times each type of possible universe $m$ in $\mathcal{M}$ is realised. The
function $f(m)$ expresses the contingency in any actualisation -- the fact
that not every possible universe has to be realised, and that any actual
universe does not have to be realised as a matter of necessity. Things could
have been different! Thus, $f(m)$ describes the \textit{ensemble of
universes } or \textit{\ multiverse} envisaged as being realised out of the
set of possibilities. If these realisations were determined by the laws of
necessity alone, they would simply be the set of possibilities described by $
\mathcal{M}$. In general they include only a subset of possible universes,
and multiple realisations of some of them. This is the way in which chance
or contingency is realised in the ensemble\footnote{It has been suggested 
to us that in mathematics terms it does not make sense
to distinguish identical copies of the same object: they should be
identified with each other because they are essentially the same. But we are
here dealing with physics rather than mathematics, and with real existence
rather than possible existence, and then multiple copies must be allowed
(for example all electrons are identical to each other; physics would be
very different if there were only one electron in existence).}.

The class of models considered is determined by all the parameters held
constant (`class parameters'). Considering the varying parameters for the
class (`member parameters'), some will take only discrete values, but for
each one allowed to take continuous values we need a volume element of the
possibility space $M$ characterised by parameter increments $dp_j(i)$ in all
such varying parameters $p_j(i)$. The volume element will be given by a
product

\begin{equation}
\pi =\Pi _{i,j}\,m_{ij}(m)\,dp_{j}(i)  \label{measure}
\end{equation}
where the product $\Pi _{i,j}$ runs over all continuously varying member
parameters $i,j$ in the possibility space, and the $m_{ij}$ weight the
contributions of the different parameter increments relative to each other.
These weights depend on the parameters $p_{j}(i)$ characterising the
universe $m$. The number of galaxies corresponding to the set of parameter
increments $dp_{j}(i)$ will be $dN$ given by

\begin{equation}
dN=f(m)\pi  \label{dist1}
\end{equation}
for continuous parameters; for discrete parameters, we add in the
contribution from all allowed parameter values. The total number of galaxies
in the ensemble will be given by 
\begin{equation}
N=\int f(m)\pi  \label{dist2}
\end{equation}
(which will often diverge), where the integral ranges over all allowed
values of the member parameters and we take it to include all relevant
discrete summations. The probable value of any specific quality $p(m)$
defined on the set of galaxies will be given by

\begin{equation}
P=\int p(m)f(m)\pi  \label{prob}
\end{equation}
Such integrals over the space of possibilities give numbers, averages, and
probabilities.

Hence, a (realised)\ ensemble $E$ of galaxies is described by a possibility
space $\mathcal{M}$, a measure $\pi $ on $\mathcal{M}$, and a distribution
function $f(m)\,$on $\mathcal{M}:$ 
\begin{equation}
E=\{\mathcal{M},\pi ,f(m)\}.  \label{ensemble}
\end{equation}

The distribution function $f(m)$ might be discrete (e. g., there are 3
copies of universe $m_1$ and 4 copies of universe $m_2$, with no copies of
any other possible universe), or continuous (e.g. characterised by a given
distribution of densities$\Omega _i$). In many cases a distribution function
will exclude many possible universes from the realisation it specifies.

Now it is conceivable that all possibilities are realised -- that all
universes in $\mathcal{M}$ exist at least once. This would mean that the
distribution function 
\[
f(m)\neq 0\mathrm{~for~all~}m\in \mathcal{M}. 
\]
But there are an infinite number of distribution functions which would
fulfil this condition, and so a really existing `ensemble of all possible
universes' is not unique. In such ensembles, all possible values of each
distinguishing parameter would be predicted to exist in different members of
the multiverse in all possible combinations with all other parameters at
least once, but they may occur many times. One of the problems is that this
often means that the integrals associated with such distribution functions
would diverge, preventing the calculation of probabilities from such models
(see our treatment of the FLRW case below).

From this consideration we have the second key issue: \newline

\noindent \textbf{Issue 2: } What determines $f(m)$? What is the meta-cause
that delimits the set of realisations out of the set of possibilities? 
\newline

The answer to this question has to be different from the answer to \textit{\
Issue 1}, precisely because here we are describing the contingency of
selection of a subset of possibilities from the set of all possibilities,
determination of the latter being what is considered in \textit{Issue 1}.
Again, the meta-question is whether this can be answered scientifically.

\subsection{Measures and Probabilities}

It is clear that $f(m)$ will enable us to derive numbers and probabilities
relative to the realisation it defines only if we also have determined a
unique measure $\pi $ on the ensemble, characterised by a specific choice of
the $\,$weights $m_{ij}(m)$ in (\ref{measure}), where these weights will
depend on the $p_{j}(i)$. There are three issues here.

\bigskip First, what may seem a \textquotedblleft natural\textquotedblright\
measure for $\mathcal{\ M}$ in one set of coordinates will not be natural in
another set of coordinates. Hence the concept of a measure is not unique, as
is illustrated below in the FLRW case. This is aggravated by the fact that
the parameter space will often contain completely different kinds of
quantities (density parameters and the values of the gravitational constant
and the cosmological constant, for example), and assigning the weights
entails somehow assigning a relative weighting between these quite different
kinds of quantities.

Second, it is possible that we might be able to assign probabilities $\chi
(m)$ to points of $\mathcal{M}$ from some kind of physical argument, and
then predict $f(m)$ from these, following the usual line of argument for
determining entropy in a gas. However, we then have to determine some reason
why $\chi (m)$ is what it is and how it then leads to $f(m)$. In the entropy
case, we assume equal probability in each phase space volume; why should
that hold for an ensemble of universes? Realising such probabilities seems
to imply a causal mechanism relating the created members of the multiverse
to one another so they are not in fact causally disjoint, otherwise, there
is no reason why any probability law (Gaussian normal, for example) should
be obeyed. We will return to this point later.

Finally, the relevant integrals may diverge. In that case, assigning mean
values or averages for physical quantities in an ensemble of universes is
problematic (see Kirchner and Ellis 2003 and references therein).

\subsection{The Anthropic subset}

The expression (\ref{life1}) can be used in conjunction with the
distribution function $f(m)$ of galaxies to determine the probability of
life arising in the whole ensemble: 
\begin{equation}
P_{life}(E)=\int f(m)*P_{gal}*R*f_S*f_p*n_e*f_l*f_i*\pi  \label{life3}
\end{equation}
(which is a particular case of (\ref{prob}) based on (\ref{life1})). An
anthropic ensemble is one for which $P_{life}(E)>0.$ If the distribution
function derives from a probability function, we may combine the probability
functions to get an overall anthropic probability function- for an example
see Weinberg et al discussed below, where it is assumed that $P_{gal}\,$ is
the only relevant parameter for the existence of life. This is equivalent to
assuming that $R*f_S*f_p*n_e*f_l*f_i=1.$ This assumption might be acceptable
in our physically realised universe, but there is no reason to believe it
would hold generally in an ensemble because these parameters will depend on
other ensemble parameters, which will vary.

\subsection{Problems With Infinity}

When speaking of multiverses or ensembles of universes -- possible or
realised -- the issue of infinity often crops up. Researchers often envision
an \textit{infinite} set of universes, in which all possibilities are
realised. Can there really be an infinite set of really existing universes?
We suggest that, on the basis of well-known philosophical arguments, the
answer is No.

There is no conceptual problem with an infinite set -- countable or
uncountable -- of \textit{possible} or \textit{conceivable } universes.
However, as stressed by David Hilbert (1964), it can be argued that a 
\textit{really existing} infinite set is not possible. As he points out,
following many others, the \ existence of the actually infinite inevitably
leads to well-recognised unresolvable contradictions in set theory, and thus
in definitions and deductive foundations of mathematics itself (Hilbert, pp.
141-142). His basic position therefore is that ``Just as operations with the
infinitely small were replaced by operations with the finite which yielded
exactly the same results . . ., so in general must deductive methods based
on the infinite be replaced by finite procedures which yield exactly the
same results.\textquotedblright\ (p. 135) He concludes, ``Our principle
result is that the infinite is nowhere to be found in reality. It neither
exists in nature nor provides a legitimate basis for rational thought . . .
The role that remains for the infinite to play is solely that of an idea . .
. which transcends all experience and which completes the concrete as a
totality . . .'' (Hilbert, p. 151). Others (see Spitzer 2000 and Stoeger 2003
and references therein) have further pointed out that realised infinite sets
are not constructible -- there is no procedure one can in principal
implement to complete such a set -- they are simply incompletable. But, if
that is the case, then ``infinity'' cannot be arrived at, or realised. On
the contrary, the concept itself implies its inability to be realised! This
is precisely why a realised past infinity in time is not considered possible
from this standpoint -- since it involves an infinite set of completed
events or moments. There is no way of constructing such a realised set, or
actualising it.

Thus, it is important to recognise that infinity is not an actual number we
can ever specify or reach -- it is simply the code-word for 
\textquotedblleft it continues without end\textquotedblright . Whenever
infinities emerge in physics -- such as in the case of singularities -- we
can be reasonably sure, as is usually recognised, that there has been a
breakdown in our models. An achieved infinity in any physical parameter
(temperature, density, spatial curvature) is almost certainly \textit{not} a
possible outcome of any physical process -- simply because it means
traversing in actuality an interval of values which never ends. We assume
space extends forever in Euclidean geometry and in many cosmological models,
but we can never prove that any realised 3-space in the real universe
continues in this way - it is an untestable concept, and the real spatial
geometry of the universe is almost certainly not Euclidean. Thus Euclidean
space is an abstraction that is probably not realised in physical practice.
In the physical universe spatial infinities can be avoided by compact
spatial sections, either resultant from positive spatial curvature or from
choice of compact topologies in universes that\ have zero or negative spatial
curvature, (for example FLRW\ flat and open universes can have finite rather
than infinite spatial sections). Future infinite time is never realised:
rather the situation is that whatever time we reach, there is always more
time available. Much the same applies to claims of a past infinity of time:\
there may be unbounded time available in the past in principle, but in what
sense can it be attained in practice? The arguments against
 an infinite past time are strong -- it's simply not constructible in terms 
of events or instants of time, besides being conceptually 
indefinite.\footnote{One way out would be, as quite a bit of work in 
quantum cosmology seems to indicate, to have time originating or emerging 
from the quantum-gravity dominated primordial substrate only ``later.'' 
In other words, there would have been a ``time'' or an epoch before time as 
such emerged. Past time would then be finite, as seems to be demanded by 
philosophical arguments, and yet the timeless primordial state could have 
lasted ``forever,'' whatever that would mean. This possibility avoids the 
problem of constructibility.}

The same problem of a realised infinity may be true in terms of the supposed
ensembles of universes. It is difficult enough conceiving of an ensemble of
many `really existing' universes that are totally causally disjoint from our
own, and how that could come into being, particularly given two important
features. Firstly, specifying the geometry of a generic universe requires an
infinite amount of information because the quantities in $\mathcal{P}_{
\mathrm{\ geometry}}$ are fields on spacetime, in general requiring
specification at each point (or equivalently, an infinite number of Fourier
coefficients) - they will almost always not be algorithmically compressible.
This greatly aggravates all the problems regarding infinity and the
ensemble. Only in highly symmetric cases, like the FLRW solutions, does this
data reduce to a finite number of parameters. One can suggest that a
statistical description would suffice, where a finite set of numbers
describe the statistics of the solution, rather than giving a full
description. Whether this suffices to adequately describe an ensemble where
`all that can happen, happens' is a moot point. We suggest not, for the
simple reason that there is no guarantee that all possible models will obey
any known statistical description. That assumption is a major restriction on
what is assumed to be possible.

Secondly, many universes in the ensemble may themselves have infinite
spatial extent and contain an infinite amount of matter, with the
paradoxical conclusions that entails (Ellis and Brundrit 1979). To conceive
of physical creation of an infinite set of universes (most requiring an
infinite amount of information for their prescription, and many of which
will themselves be spatially infinite) is at least an order of magnitude
more difficult than specifying an existent infinitude of finitely
specifiable objects.

The phrase `everything that can exist, exists' implies such an infinitude,
but glosses over all the profound difficulties implied. One should note here
particularly that problems arise in this context in terms of the continuum
assigned by classical theories to physical quantities and indeed to
spacetime itself. Suppose for example that we identify corresponding times
in the models in an ensemble and then assume that \textit{all} values of the
density parameter occur at each spatial point at that time. Because of the
real number continuum, this is an uncountably infinite set of models -- and
genuine existence of such an uncountable infinitude is highly problematic.
But on the other hand, if the set of realised models is either finite or
countably infinite, then almost all possible models are not realised -- the
ensemble represents a set of measure zero in the set of possible universes.
Either way the situation is distinctly uncomfortable. However we might try
to argue around this by a discretization argument:\ maybe differences in
some parameter of less than say $10^{-10}$ are unobservable, so we can
replace the continuum version by a discretised one, and perhaps some such
discretisation is forced on us by quantum theory. If this is the intention,
then that should be made explicit. That solves the `ultraviolet divergence'
associated with the small-scale continuum, but not the `infrared divergence'
associated with supposed infinite distances, infinite times, and infinite
values of parameters describing cosmologies.

\section{Ensembles of FLRW Universes}

Having established the broad set of issues concerning multiverses that we
believe need to be addressed, we shall for the remainder of this paper limit
ourselves to the FLRW sector $\mathcal{M}_{\mathrm{FLRW}}$ of the ensemble
of all possible universes $\mathcal{M}$ in order to illustrate these 
issues\footnote{Many discussions implicitly suggest that this is the 
whole possibility space,
as they only consider FLRW\ models as possibilities. However these clearly
form a very small subspace of all geometrical possibilities.}. We assume the
family considered is filled with matter components characterised by a $
\gamma $-law equation of state, and mainly restrict our attention to their
cosmological parameters, although full consideration of anthropic issues
would be characterised by including all the other parameters. Our
descriptive treatment will consider FLRW universe domains (whether a true
multiverse or separate domains in a single spacetime) as distinct but with
common physical characteristics.

\subsection{Properties of FLRW models}

FLRW models are homogeneous and isotropic models described by the metric 
\begin{equation}
\mathrm{d}s^2=-\mathrm{d}t^2+a^2(t)\left( \frac{{}r^2}{1-kr^2}+r^2\mathrm{d}
\Omega ^2\right) ,  \label{metric}
\end{equation}
where $\mathrm{d}\Omega ^2=\mathrm{d}\vartheta ^2+\sin ^2(\vartheta )\mathrm{
\ \ \ \ \ d}\varphi ^2$ denotes the line element on the two-dimensional unit
sphere, $a(t)$ is the scale-factor, and 
\[
k=\left\{ 
\begin{array}{ll}
1 & \mbox{for closed models} \\ 
0 & \mbox{for flat models} \\ 
-1 & \mbox{for open models}
\end{array}
\right. 
\]
is the normalised curvature. The FLRW model is completely determined by $k$
and the scale-factor $a(t)$, which incorporates the time-evolution and is
obtained from the Einstein-Field equations together with the matter
description.\footnote{The way the tetrad description given above 
relates to FLRW universes is
described in detail in Ellis and MacCallum\ (1968); the standard coordinates
given here are more convenient if one discusses only the FLRW\ models.}

Assuming gravity is described by the Einstein field equations, the evolution
of FLRW models is described by the Friedmann equation 
\begin{equation}
H^2(\Omega -1)=\frac k{a^2},  \label{Fried}
\end{equation}
where $H \equiv \dot{a}/a$ (a dot denotes differentiation with respect to
proper time) is the Hubble parameter and $\Omega $ the density parameter. We
restrict our discussion to models with only a cosmological constant $\Lambda 
$ and one matter component which obeys a $\gamma $-law equation of state,
i.e., its pressure $p$ and density $\rho $ are related by $p=(\gamma -1)\rho 
$, where $\gamma $ is constant. This specification of parameters $p_3(i)$
includes in particular the case of dust ($\gamma =1$) and radiation ($\gamma
=4/3$). The total density parameter is 
\begin{equation}
\Omega =\Omega _m + \Omega _\Lambda .  \label{omega}
\end{equation}
where the matter density parameter is $\Omega_m \equiv \frac{\kappa \rho }{
3H^2}$ and the vacuum-energy density parameter is $\Omega _\Lambda \equiv
\frac \Lambda {3H^2}$ (representing a cosmological constant). These form the
parameters $p_4(i).$

The second time derivative of the scale factor is determined by the
Raychaudhuri equation 
\begin{equation}
2q=(3\gamma -2)\Omega _m-\Omega _\Lambda ,  \label{ray}
\end{equation}
where $q\equiv -\frac{\ddot{a}}{aH^2}$ is the dimensionless deceleration
parameter. The matter evolution is given by the energy-conservation equation 
\begin{equation}
\dot{\rho}=-3\gamma \rho H  \label{cons}
\end{equation}
or equivalently by 
\begin{equation}
\dot{\Omega}_m=\Omega _mH(q+1-3\gamma ).  \label{omega_dot}
\end{equation}
Besides the normalised curvature $k$ there are two constants of motion, $
\chi \equiv \kappa \rho a^{3\gamma }/3$ and the cosmological constant $
\Lambda $. Given these parameters, the dynamical evolution is determined
from the implied initial conditions: \{$a(t_0),\Omega _{i0},\gamma
_i,k\}\Rightarrow a(t)$.

\subsection{Parametrising FLRW models}

In order to define a FLRW ensemble we need a set of independent parameters
which uniquely identify all possible models. We want to consider all
possible FLRW models with the same physical laws as in our universe, but
possibly different coupling constants. There is then one set of parameters $
p_2(i)$ which defines the gravitational ``physics'' of the model in terms of
the coupling constants -- for simplicity let us only consider the
gravitational constant $G$ here -- and further sets $p_5(i),$ $p_4(i)$ which
identify the geometry and matter content of the actual model, and which are
related to each other via the Einstein field equations.

Among the various options there are two particularly useful
parametrisations. Ehlers and Rindler (1989) developed a parametrisation in
terms of the observable density parameters (they also include a radiation
component) and the Hubble parameter. With $\Omega _k \equiv \frac k{H^2a^2}$
the Friedmann equation becomes 
\begin{equation}
\Omega_m + \Omega _\Lambda -1=\Omega _k.  \label{uli-friedmann}
\end{equation}
The curvature parameter $\Omega _k$ determines $k = sgn(\Omega _k).\,$For $k
\neq 0$ the scale-factor, and hence the metric (\ref{metric}), is determined
by 
\[
a^2(t)=\frac k{H^2\Omega _k}=\frac k{H^2(\Omega_m + \Omega_\Lambda -1)}, 
\]
while for $k=0$ its value is unimportant because of scale-invariance in that
case. Hence any \emph{state} is completely described by $\Omega_m, \Omega
_\Lambda $, and $H$.

In order to parametrise the models rather than the states, we need to select
one particular time $t_0$ for each model at which we take the above
parameters as representative parameters $\Omega_{m 0}, \Omega_{\Lambda 0}$,
and $H_0$ for this model.\footnote{Hence, in general different models 
correspond to the same values of $
\Omega_m $ and $\Omega_\Lambda$, depending on the value of $H$.} We note
that this time $t_0$ can be model dependent because not all models will
reach the age $t_0$.

All big-bang FLRW models start with an infinite positive Hubble parameter
whose absolute value reaches or approaches asymptotically a minimum value $
H_{\mathrm{min}}$. Hence we could define the time $t_{0}$ as the time when
the model first takes a certain value $H_{0}(p_{I})>H_{\mathrm{min}}(p_{I})$
, where $p_{I}$ represents the model parameters. One particular choice of $
H_{0}(p_{I})$ is given by 
\[
H_{0}(p_{I})\equiv \exp (H_{\mathrm{min}}(p_{I})) 
\]
On the other hand, by setting $H_{0}(p_{I})=\mathrm{const}$ and excluding
all models which never reach this value one finds easily a parametrisation
of all models which reach this Hubble value during their evolution.

While above choice of parameters give a convenient parametrisation in terms
of observables which covers closed, flat, and open models, it is disturbing
that for each model an arbitrary time has to be chosen. This also leads to a
technical difficulty, because the parameters $\{H_{0},\Omega
_{m0},\Omega_{\Lambda 0}\}$ are subject to the constraint $
H_{0}=H_{0}(p_{I}) $.

For these reasons it is often convenient to use a set of parameters which
are comoving in the state-space, i.e., parameters which are constants of
motion. As mentioned above, for open and closed models such a set is given
by the matter constant $\chi$, the cosmological constant $\Lambda$, and the
normalised curvature constant $k$. For flat models one can rescale the scale
factor, which allows us to set $\chi=1$.

These parameters are related to the observational quantities by (for $k=\pm1$
) 
\[
\chi =\frac{\Omega _{m}H^{2}}{(H^{2}|1-\Omega _{m}-\Omega _{\Lambda
}|)^{3\gamma /2}}\quad \mbox{ and }\quad \Lambda =3H^{2}\Omega _{\Lambda }. 
\]

The evolution of these models through state space is illustrated here in
terms of two different parametrisations of the state space, see Figures 1a
and 1b. For a detailed investigation of these evolutions for models with
non-interacting matter and radiation, see Ehlers and Rindler (1989).

\begin{figure*}[tbp]
\centering          
\subfigure[$\Omega_m-H$]{\ \label{flrw-phase-space-a} 
\includegraphics[width=3in]{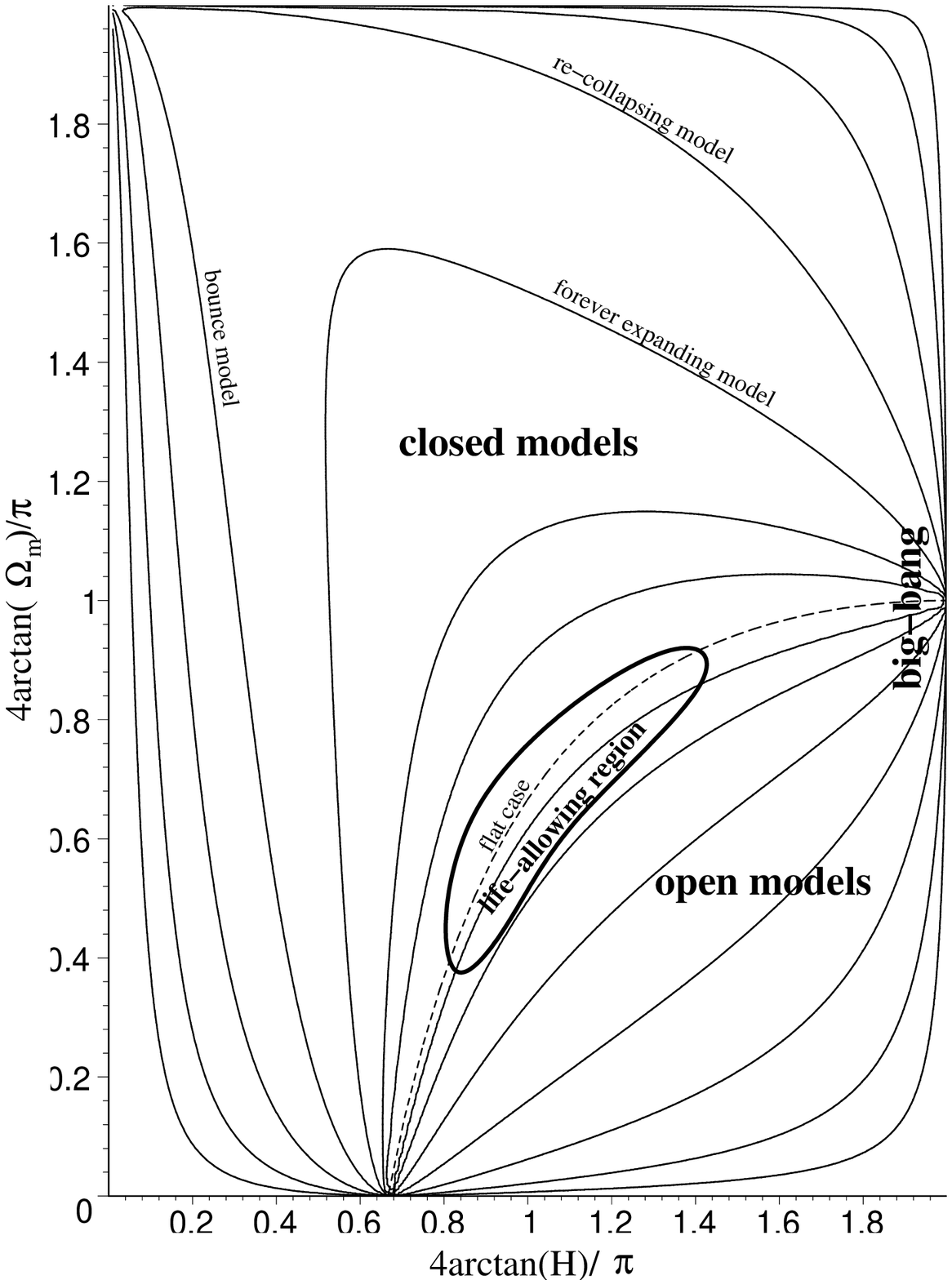}} \hspace{.5cm} \subfigure[$\Omega_
\Lambda-\Omega_m$- plane]{\includegraphics[width=3in]{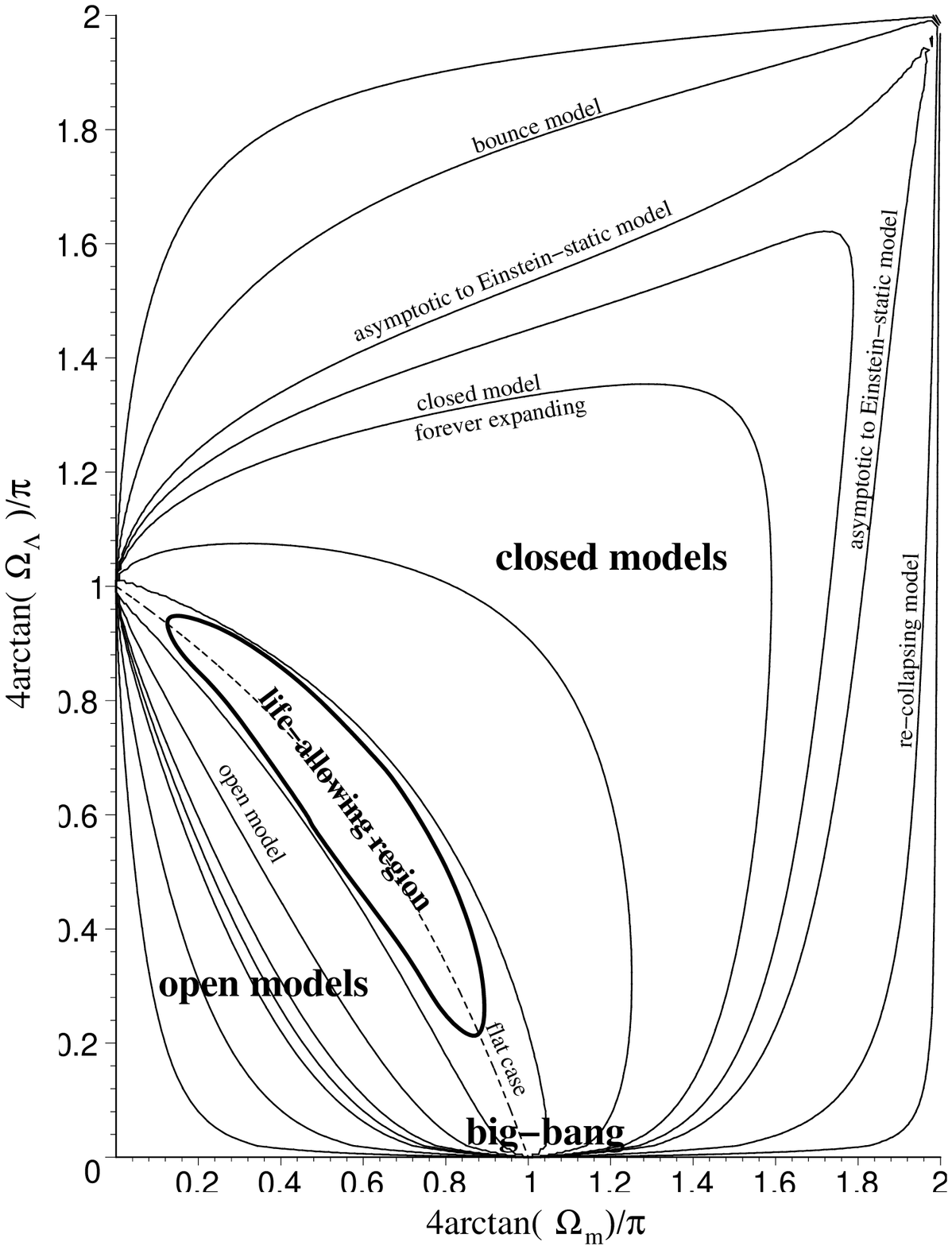}}
\caption{Phase-space diagrams showing the evolution of FLRW-models (as given
by the Friedmann equation) in the $\Omega_m-H$ and $\Omega_\Lambda-\Omega_m$
-plane for $\Lambda=1$. Models can evolve in both directions along the
lines. Models which start with a big-bang originate on the right in figure
(a) (corresponding to $H=\infty$). Models which reach $H=0$ (in a finite
time) reverse their direction of evolution, i.e., they follow the same line
back-wards. These are the re-collapsing models. The limiting case to the
forever-expanding models is given by the model which approaches
asymptotically the Einstein-static universe.}
\label{omega-H-rel}
\end{figure*}

\subsection{The possibility space}

The structures defined so far are the uniform structures across the class of
models in this possibility space, characterised both by laws of physics (in
particular General Relativity) and by a restricted class of geometries. It
is clear that universes in a multiverse should be able to differ in at least
some properties from each other. We have just characterised the geometrical
possibilities we are considering. The next question is, which physical laws
and parameters can vary within the ensemble, and which values can they take?
For this simplified discussion let us just assume that only the
gravitational constant $G$ and the cosmological constant $\Lambda $ (which
also qualifies as a model parameter) are variables, with the 
ranges\footnote{It is worth noting that when $G=0$, 
we \emph{do} obtain FLRW solutions to
the Einstein-Field equations: the Milne universe which is effectively empty
-- no gravity effective.} $G\in [0,\infty )$ and $\Lambda \in (-\infty
,\infty )$. However, if we consider \textquotedblleft all that is possible
\textquotedblright\ within this restricted class of FLRW\ models, maybe we
should consider $G\in (-\infty ,\infty ).$ There is still considerable
uncertainty as to the nature of an ensemble even within this restricted
context. Whatever is chosen here defines the set of possibilities that can
arise.

\subsection{The measure}

For a complete probabilistic description of an ensemble we need not only a
distribution function $P$, but also a measure $\pi$ for the parameter
space (see Section 3.1). The information entropy 
\begin{equation}
S\equiv -\int \mathrm{d}xP(x)\log \left( \frac{P(x)}{\mu (x)}\right)
\label{uli-entropy}
\end{equation}
is then maximised for the probability distribution equal to this measure,
representing the state of minimal knowledge.

Without knowledge of the creation mechanism it is impossible to determine
this measure with certainty. Nevertheless, we might ask what our best guess
for such a measure should be in a state of minimal information, where only a
certain set of independent parameters, describing the ensemble, and their
ranges are known.

The only known method for constructing such a measure is Jaynes' principle.
Its application to FLRW models with $\gamma$-law equation of state has been
discussed in Kirchner and Ellis (2003). One identifies a set of
transformations $x^{\prime }(x)$ in possibility space which leaves the
mathematical structure invariant, and demands that the measure is invariant
under these transformations. The two most important cases are given by
parameters which can take all real values, and those that take on all
positive real values.

In the first case, if $z$ is a valid parameter value, so is $z^{\prime
}=z+\alpha $ for all real $\alpha $. According to Jaynes' principle the
measure should obey $\mathrm{\pi =}\mu (z)\mathrm{d}z=\mu (z^{\prime })
\mathrm{d}z^{\prime }$ and hence $\mu (z)=\mu (z^{\prime })$, i.e., the
measure is constant.

If on the other hand a parameter $u$ only takes non-zero positive values, we
can generate another valid parameter value by $u^{\prime }=\lambda u$, where 
$\lambda \in R^{+}$. Demanding invariance of the measure yields $\mathrm{\pi 
}\propto \frac 1u \mathrm{d}u$.

These transformations are not unique, and hence one could find many
different measures. Nevertheless, in the state of minimum information we
don't know what the natural parametrisation is for the possibility space and
different measures correspond to different guesses. Surprisingly Jaynes'
principle is ``relatively invariant'' under simple parametrisation changes.
For example, introducing a new parametrisation for a positive quantity $G$
by $G = m^{n}$ for positive $m$, or by $G = \exp (\lambda )$ for real $
\lambda $, will give the same measure.

There are two important points to note. Firstly the measure is derived from
the chosen set of parameters. Generally a different choice of parameters
yields a different minimum-information measure, predicting another
maximum-entropy distribution function. Let us consider the example of an
ensemble of dust-FLRW\ models. The different open and closed models are most
conveniently parametrised by the constants of motion, which are given by the
cosmological constant $\Lambda $ and $\chi \equiv a\rho ^{3\gamma }$, where $
\rho $ is the energy density. This leads to the minimum-information measure
(Kirchner and Ellis 2003). 
\begin{equation}
\mathrm{\pi }\propto \frac{\mathrm{d}\Lambda \mathrm{d}\chi }{\sqrt{\chi }}.
\label{uli-measure-1}
\end{equation}
Considering dust models ($\gamma =1$) and the subset of all big-bang models
which reach a certain Hubble parameter $H_0$ at a time $t_0$ during their
evolution this measure becomes 
\[
\mathrm{\pi }\propto \sqrt{\frac{\Omega _{m0}}{|\Omega _0-1|^{3/2}}}\left|
\frac 1{\Omega _{m0}}-\frac{3/2}{\Omega _0-1}\right| \mathrm{d}\Omega _{m0}
\mathrm{d}\Omega _{\Lambda 0},
\]
with $\Omega _{\Lambda 0}\leq 1+\Omega _{m0}/2$. On the other hand, as
mentioned above, there is a convenient parametrisation for this particular
subset of models (Ehlers and Rindler 1989) in terms of the observables $
\Omega _{m0}$ and $\Omega _{\Lambda 0}$ (in Ehlers and Rindler (1989) an
additional radiation component was also included). Using this
parametrisation yields the minimum information measure 
\[
\mathrm{\pi }\propto \frac 1{\sqrt{\Omega _{m0}}}\mathrm{d}\Omega _{m0}
\mathrm{d}\Omega _{\Lambda 0},
\]
which is clearly different from the above result.

Secondly, the measure is in general non-normalisable and hence there is no
normalisable maximum-entropy distribution. Without additional information we
are not able to calculate certain probabilities. Since it seems questionable
whether there will ever be additional information about the ensemble of
universes available, one has to accept that certain questions will have no
well defined probabilities.

It should be mentioned that we encounter similar problems when we want to
find a probability measure for physical parameters like the gravitational
constant $G$. Let us assume that $G$ can take any non-zero positive value.
Jaynes' principle then suggests the probability measure $\mathrm{\pi }
_G\propto \frac{\mathrm{d}G}G$. On the other hand, if we decide to use 
\footnote{However, it should be noted that 
any power ($m=G^n,m\in {^+},n\neq 0$) and
logarithmic relationship ($m=\ln (G),m\in $) leads to the same measure $
\mathrm{d}\mu _{G}$.} $m=\sinh (G)$ as our parameter, then we find the
different measure 
\[
\mathrm{\pi =}\frac{\mathrm{d}m}m=\frac{\cosh (G)}{\sinh (G)}\mathrm{d}G.
\]

\subsection{Distribution Functions on $\mathcal{M}_\mathrm{FLRW}$}

Now, having properly parametrised $\mathcal{M}_{\mathrm{FLRW}}$ and defined
a measure on it, we can represent particular multiverses by giving
distribution functions over the parameter-space (as discussed in Section
3.4). Given a distribution function $f$ it determines the number of
universes in a small parameter-interval by 
\[
\mathrm{d}N=f(p_I)\mathrm{\pi },
\]
which is invariant under a change of parametrisation. Hence it is the
combination of measure and distribution function which is of importance.

While distribution functions can be parametrised by any set of coordinates
over the possibility space, we need different distribution functions for
different possibility spaces. For example, if universes in a multiverse 
\emph{must} have a common value for the gravitational constant $G$ then
distribution functions must not depend on $G$.

It is clear that a particular distribution function can be expressed in any
set of coordinates. Obviously there is a vast set of possible distribution
functions. We want to examine some particular examples.

Firstly, one could have a distribution function which is constant over the
parameter-space. The actual ensemble then really depends on the measure and
is the maximum-entropy distribution (which maximises 
(\ref{uli-entropy})\footnote{In (\ref{uli-entropy}) $P(x) \mathrm{d}x$ 
corresponds to $f(p_I)\pi$.}). 
If we choose the observational quantities $H,\Omega _m,\Omega
_\Lambda $ to represent the model and allow for different values of the
gravitational coupling constant $G$ this would be 
\[
f(H,\Omega _m,\Omega _\Lambda ,G)=\mathrm{const.}
\]
for all allowed values of the stated parameters. On the other hand, if we
choose the constants of motion as coordinates in possibility space 
\[
f(k,\chi ,\Lambda ,G)=\mathrm{const.}
\]
The probability $P_{\mathcal{A}}$ to find a universe in a certain
parameter-region $\mathcal{A}$ is given by 
\[
P_{\mathcal{A}}=\frac{\int_{\mathcal{A}}f(k,\chi ,\Lambda ,G)\mathrm{\pi }.}{
\int f(k,\chi ,\Lambda ,G)\mathrm{\pi }.},
\]
where the integral in the denominator extends over the whole possibility
space. For many distribution functions, like for the above constant
distribution function together with (\ref{uli-measure-1}), this expression
is not well defined. Let us assume that the measure is non-integrable, i.e.,
non-normalisable, and that we have a constant distribution function. If the
set $\mathcal{A}$ does not include any point of non-integrability of the
measure then $P_{\mathcal{A}}=0$, if it includes all points of
non-integrability then $P_{\mathcal{A}}=1$, but if it includes only some of
the non-integrabilities then $P_{\mathcal{A}}$ is not well defined.

Of course, the above expression might be integrable for all sets $\mathcal{A}
$ given a good distribution function. For instance the distribution function 
\[
f(k, \chi,\Lambda )\propto \exp (-\chi -\Lambda ^{2}) 
\]
together with the measure (\ref{uli-measure-1}) is integrable everywhere. On
the other hand 
\[
f(k, \chi ,\Lambda )\propto \frac{\exp (-\Lambda ^{2})}{\sqrt{\chi +1}} 
\]
diverges for $\chi \rightarrow \infty $. A distribution function can also
introduce an additional divergence, for instance 
\[
f(k, \chi, \Lambda) \propto \frac{\exp (-\Lambda ^{2})}{\sqrt{\chi }} 
\]
is non-integrable at $\chi =0$ and $\chi \rightarrow \infty $.

A multiverse might contain a finite or infinite countable number of
universes. In these cases the distribution function contains Dirac\ $\delta$
-functions, e.g., 
\[
f(k, \chi, \Lambda)= \left\{ \matrix{\ \sqrt{\chi}5\delta^2(\chi-3,
\Lambda-.5) & \mbox{ for
}k=1\cr
2\delta(\Lambda-.7)& \mbox{ for }k=0\cr
\sqrt{\chi}3\delta^2(\chi-.1,\Lambda-2)& \mbox{ for }k=-1 } \right. 
\]
which represents a multiverse which contains $5$ copies of closed FLRW
models with $\chi=3$ and $\Lambda=.5$, etc. The distribution function 
\[
f(k, \chi, \Lambda)= \left\{ \matrix{\ \sqrt{\chi} \sum_{i=1}^\infty
\delta(\chi-1/i, \Lambda-1) & \mbox{ for } k=1\cr 0 &\mbox{
for } k \ne 1 } \right. 
\]
represents an ensemble with a countably infinite number of universes -- all
are closed with $\Lambda=1$ and one for each $\chi=1/i$. Similarly one could
imagine an ensemble of $10^7$ copies of our universe, which would be
represented by the distribution function 
\[
f(k,\chi, \Lambda) = 10^7 \sqrt{\chi} \delta^{k_0}_{k}
\delta(\chi-\chi_0)\delta(\Lambda-\Lambda_0), 
\]
where $k_0, \chi_0, \Lambda_0$ represent the parameter values for our
universe. This is very unlikely in terms of a generating mechanism, but for
ensembles without generating mechanisms it is as likely as any other
possibility. If a multiverse is ``tested'' by its prediction that our
universe is a likely member, then such an ensemble should be the most
satisfying one -- but then we might just as well be happy with one copy,
i.e., just our universe.

Similar distribution functions determine the distribution of physical
parameters like the gravitational constant $G$. For example with $G \in 
\mathbf{R}$ the minimum information measure is $\mathrm{d}\mu _G=\mathrm{d}G$
and 
\[
f(G)=\exp (-G^2) 
\]
gives a Gaussian distribution around $G=0$. If, on the other hand $G\in
(0,\infty )$ the measure is $\mathrm{d}G/G$ and $f(g)=G\exp (-G)$ would be
an example of a distribution function.

One can imagine various types of distributions, e. g., a Gaussian
distribution in $G$ or in $H_{0}$, or in the other parameters. But, in order
to establish these in a non-arbitrary way, we need a theory of how this
particular ensemble is selected for from all the other possible ones.

\subsection{The problem of infinities again}

Even within the restricted set of FLRW\ models, one of the most profound
issues is the problem of realised infinities: if all that is possible in
this restricted subset happens, we have multiple infinities of realised
universes in the ensemble. First, there are an infinite number of possible
spatial topologies in the negative curvature case, so an infinite number of
ways that universes which are locally equivalent can differ globally.
Second, even though the geometry is so simple, the uncountable continuum of
numbers plays a devastating role locally: is it really conceivable that FLRW
universes actually occur with \textit{all} values independently of both the
cosmological constant and the gravitational constant, and also all values of
the Hubble constant at the instant when the density parameter takes the
value 0.97? This gives 3 separate uncountably infinite aspects of the
ensemble of universes that is supposed to exist. The problem would be
allayed if spacetime is quantized at the Planck level, as suggested for
example by loop quantum gravity. In that case one can argue that all
physical quantities also are quantized, and the uncountable infinities of
the real line get transmuted into finite numbers in any finite interval -- a
much better situation. We believe that this is a physically reasonable
assumption to make, thus softening a major problem for many ensemble
proposals. But the intervals are still infinite for many parameters in the
possibility space. Reducing the uncountably infinite to countably infinite
does not in the end resolve the problem of infinities in these ensembles. It
is still an extraordinarily extravagant proposal.

\subsection{The anthropic subset}

We can identify those FLRW universes in which the emergence and sustenance
of life is possible at a broad level\footnote{More accurately, 
perturbations of these models can allow life -- the exact
FLRW models themselves cannot do so.} -- the necessary cosmological
conditions have been fulfilled allowing existence of galaxies, stars, and
planets if the universe is perturbed, so allowing a non-zero factor $\Pi =$ $
P_{gal}*R*f_S*f_p*n_e\,$ as discussed above. These are indicated in the
Figures above (anthropic universes are those intersecting the regions
labelled ``life allowing''). The fraction of these that will actually be
life-bearing depends on the fulfilment of a large number of other conditions
represented by the factor $F=f_l*f_i,\,$ which will also vary across a
generic ensemble, and the above assumes this factor is non-zero.$.$

\section{On the origin of ensembles}

Ensembles have been envisaged both as resulting from a single causal
process, and as simply consisting of discrete entities. We discuss these two
cases in turn, and then show that they are ultimately not distinguishable
from each other.

\subsection{Processes Naturally Producing Ensembles}

Over the past 15 or 20 years, many researchers investigating the very early
universe have proposed processes at or near the Planck era which would
generate a really existing ensemble of expanding universe domains, one of
which is our own observable universe. In fact, their work has provided both
the context and stimulus for our discussions in this paper. Each of these
processes essentially selects a really existing ensemble through a
generating process from a set of possible universes, and often lead to
proposals for a natural definition of a probability distribution on the
space of possible universes. Here we briefly describe some of these
proposals, and comment on how they fit within the framework we have been
discussing.

Andrei Linde's (1983, 1990) chaotic inflationary proposal (see also Linde
(2003) and references therein) is one of the best known scenarios of this
type. The scalar field (inflaton) in these scenarios drives inflation and
leads to the generation of a large number of causally disconnected regions
of the Universe. This process is capable of generating a really existing
ensemble of expanding FLRW-like regions, one of which may be our own
observable universe region, situated in a much larger universe that is
inhomogeneous on the largest scales. No FLRW\ approximation is possible
globally; rather there are many FLRW-like sub-domains of a single fractal
universe. These domains can be very different from one another, and can be
modelled locally by FLRW cosmologies with different parameters.

Linde and others have applied a stochastic approach to inflation
(Starobinsky 1986, Linde, \textit{et al.} 1994, Vilenkin 1995, Garriga and
Vilenkin 2001, Linde 2003), through which probability distributions can be
derived from inflaton potentials along with the usual cosmological equations
(the Friedmann equation and the Klein-Gordon equation for the inflaton) and
the slow-roll approximation for the inflationary era. A detailed example of
this approach, in which specific probability distributions are derived from
a Langevin-type equation describing the stochastic behaviour of the inflaton
over horizon-sized regions before inflation begins, is given in Linde and
Mezhlumian (2003) and in Linde \textit{et al.} (1994). The probability
distributions determined in this way generally are functions of the inflaton
potential.

This kind of scenario suggests how overarching physics, or a 
\textquotedblleft law of laws\textquotedblright (represented by the inflaton
field and its potential), can lead to a really existing ensemble of many
very different FLRW-like regions of a larger Universe. However these
proposals rely on extrapolations of presently known physics to realms far
beyond where its reliability is assured. They also employ inflaton
potentials which as yet have no connection to the particle physics we know
at lower energies. And these proposals are not directly observationally
testable -- we have no astronomical evidence the supposed other FLRW-like
regions exist. Thus they remain theoretically based proposals rather than
established fact. There additionally remains the difficult problem of
infinities: eternal inflation with its continual reproduction of different
inflating domains of the Universe is claimed to lead to an infinite number
of universes of each particular type (Linde, private communication). How can
one deal with these infinities in terms of distribution functions and an
adequate measure? As we have pointed out above, there is a philosophical
problem surrounding a realised infinite set of any kind.

Finally, from the point of view of the ensemble of all possible universes
often invoked in discussions of multiverses, all possible inflaton
potentials should be considered, as well as all solutions to all those
potentials. They should all exist in such a multiverse, which will include
chaotic inflationary models which are stationary as well as those which are
non-stationary. Many of these potentials may yield ensembles which are
uninteresting as far as the emergence of life is concerned, but some will be
bio-friendly. The price of this process for creating anthropically
favourable universe regions is the multiplication of realised infinities,
most of which will be uncountable (for example the parameters in any
particular form of inflaton potential will take all possible values in an
interval of real numbers).

\subsection{Probability distributions for the cosmological constant}

Weinberg (2000) and Garriga and Vilenkin (2001) derive a probability
distribution for the cosmological constant in the context of an ensemble of
regions generated in the same inflationary sequence via the action of a
given inflaton potential where the cosmological constant is given by the
potential energy of a scalar-field. In multi-domain universes, where spatial
variations in a scalar-field cause different regions to inflate at different
rates, the cosmological constant should be distributed according to some
probability distribution $P(\rho _\Lambda )$. During inflation the scalar
field undergoes randomisation by quantum fluctuations, such that later on
its values in different regions are distributed according to 
\textquotedblleft the length\textquotedblright\ in field space (Garriga and
Vilenkin 2002). This leads to a probability distribution (or distribution
function -- the probability distribution is just the normalised distribution
function) of values of the vacuum-energy density $\rho _\Lambda $ in these
regions given by 
\[
P(\rho _\Lambda )\mathrm{d}\rho _\Lambda \propto \frac{\mathrm{d}\rho
_\Lambda }{\mid V^{\prime }(\phi )\mid }, 
\]
where $V(\phi )$ is the inflaton potential, and the prime signifies
differentiation with respect to the inflaton $\phi $.

It has been suggested (Vilenkin 1995, Weinberg 1997) that the way the
probability distribution for existence of galaxies depends on the
cosmological constant can be approximated by

\begin{equation}
P_{gal}=N(\rho _\Lambda )P(\rho _\Lambda )  \label{dm-l-01}
\end{equation}
where $N(\rho _\Lambda )$ is the fraction of baryons that form galaxies. The
requirement of structure formation as a pre-requisite for life places strong
anthropic constraints on the domains in which observers could exist; these
constraints must be satisfied in the really existing universe.

Let us first note that galaxy formation is only possible for a narrow range
around $\rho _\Lambda =0$ (Weinberg 2000). It has been shown that anthropic
restrictions demand $\rho _\Lambda \lesssim 10^{-28} \rm \frac g{cm^3}$ (Kallosh
and Linde 2002, Garriga and Vilenkin 2002). Consequently the anthropic
selection factor $N(\rho _\Lambda )$ is sharply peaked and vanishes for $
|\rho _\Lambda |>\rho _{\Lambda \max }$ for some $\rho _{\Lambda \max }$,
which is of the same order of magnitude as the observed cosmological
constant. In scalar-field models $P(\rho _\Lambda )$ is in direct relation
to the \textit{a priori} distribution of the scalar-field fluctuations and
it has been argued (Weinberg 2000) that for a wide class of potentials the
variations of $P(\rho _\Lambda )$ over the anthropically allowed range
(where $N(\rho _\Lambda )\neq 0$) should be negligible. Nevertheless, as has
been shown in (Vilenkin-Garriga 2002) this is not always the case, in
particular for power-law potentials $V(\phi )=\phi ^n$ with $n>1$ one finds
an integrable divergence at $\rho _\Lambda =0$.

It is clear that a similar relation to (\ref{dm-l-01}) should hold for
multiverses in the wider sense. Nevertheless, one could imagine multiverses
containing universes with and without scalar-field, or with different
potentials. Hence we cannot link the distribution of the cosmological
constant to that of the scalar-field in a unique way, and there is a vast
choice for possible \textit{a priori }probability distributions for the
cosmological constant. Let us assume that the cosmological constant is a
remnant of some underlying (unknown) theory and as such might be restricted
to some domain of values. Depending on this domain one finds different
possible minimum information measures, which result from Jaynes' principle.
If the domain is given by all real numbers then the (non-normalisable)
measure will be constant. If on the other hand the domain is given by all
positive real numbers then the minimum-information measure gives an
(non-normalisable) \textit{a priori} probability distribution proportional
to $1/\rho _\Lambda $ (Kirchner and Ellis 2003). In this case the divergence
is located inside the anthropically allowed region and is non-integrable.
For this case the expectation value vanishes, i.e., 
\[
\bar{\rho}_\Lambda =\frac{\int_0^{\rho_{\Lambda \max}}\rho _{\Lambda} \frac 1{\rho
_\Lambda }\mathrm{d} \Lambda }{\int_0^{\rho_{\Lambda \max }} \frac 1{\rho _\Lambda
}\mathrm{d}\Lambda }=0 
\]
and we fail to explain the observed non-zero value of the cosmological
constant.

An interesting alternative is given by allowing the cosmological constant to
take values in the domain $R^{+}\cup \{0\}$ (e.g., if the cosmological
constant prediction is given by a quadratic term). The minimum
information-measure is then proportional to $1/\sqrt{\rho _\Lambda }$. Again
there is a divergence in the anthropically allowed region, but this time it
is integrable. The expectation value becomes 
\[
\bar{\rho}_\Lambda =\frac{\int_0^{\rho _{\Lambda \max }}\rho _\Lambda \frac 1{
\sqrt{\rho _\Lambda }}\mathrm{d}\Lambda }{\int_0^{\rho _{\Lambda \max }}\frac 1{
\sqrt{\rho _\Lambda }}\mathrm{d}\Lambda }=\frac 13\rho _{\Lambda \max }. 
\]

\subsection{The existence of regularities}

Consider now a genuine multiverse. Why should there be any regularity at all
in the properties of universes in such an ensemble, where the universes are
completely disconnected from each other? If there are such regularities and
specific resulting properties, this suggests a mechanism creating that
family of universes, and hence a causal link to a higher domain which is the
seat of processes leading to these regularities. This in turn means that the
individual universes making up the ensemble are not actually independent of
each other. They are, instead, products of a single process, as in the case
of chaotic inflation. A common generating mechanism is clearly a causal
connection, even if not situated in a single connected spacetime -- and some
such mechanism is needed if all the universes in an ensemble have the same
class of properties, for example being governed by the same physical laws or
meta-laws.

The point then is that, as emphasized when we considered how one can
describe ensembles, any multiverse with regular properties that we can
characterise systematically is necessarily of this kind. If it did not have
regularities of properties across the class of universes included in the
ensemble, we could not even describe it, much less calculate any properties
or even characterise a distribution function.

Thus in the end the idea of a completely disconnected multiverse with
regular properties but without a common causal mechanism of some kind is not
viable. There must necessarily be some pre-realisation causal mechanism at
work determining the properties of the universes in the ensemble. What are
claimed to be totally disjoint universes must in some sense indeed be
causally connected together, albeit in some pre-physics or meta-physical
domain that is causally effective in determining the common properties of
the multiverse.

Related to this is the issue that we have emphasized above, namely where
does the possibility space come from and where does the distribution
function come from that characterises realised models? As emphasized above,
we have to assume that some relevant meta-laws pre-exist. We now see that we
need to explain also what particular meta-laws pre-exist. If we are to
examine `all that might be, exists', then we need to look at the ensemble of
all such meta-laws and a distribution function on this set. We seem to face
an infinite regress as we follow this logic to its conclusion, and it is not
clear how to end it except by arbitrarily calling a stop to this process.
But then we have not looked at all conceivable possibilities.

\section{Testability and Existence}

Finally, the issue of evidence and testing has already been briefly
mentioned. This is at the heart of whether an ensemble or multiverse
proposal should be regarded as physics or as metaphysics.

\subsection{Evidence and existence}

Given all the possibilities discussed here, which specific kind of ensemble
is claimed to exist? Given a specific such claim, how can one show that this
is the particular ensemble that exists rather than all the other
possibilities?

There is no direct evidence of existence of the claimed other universe
regions, nor can there be any, for they lie beyond the visual horizon; most
will even be beyond the particle horizon, so there is no causal connection
with them; and in the case of a true multiverse, there is not even any
possibility of any indirect causal connection of any kind - the universes
are then completely disjoint and nothing that happens in any one of them is
linked to what happens in any other one.

What weight does a claim of such existence carry, in this context when no
direct observational evidence can ever be available? The point is that there
is not just an issue of showing a multiverse exists - if this is a
scientific proposition one needs to be able to show which specific
multiverse exists; but there is no observational way to do this. Indeed if
you can't show which particular one exists, it is doubtful you have shown
any one exists. What does a claim for such existence mean in this context?

These issues are discussed in more depth in the accompanying philosophical
paper, where we consider the various ways one may claim entities exist even
when there is no direct or even indirect evidence for such existence. One
ends up in deep philosophical waters. That is unavoidable if one is to
seriously argue the claim for existence of a multiverse. Even the concept of
what `existence' might mean in this context needs careful consideration.

\subsection{Observations and Physics}

The one way one might make a reasonable claim for existence of a multiverse
would be if one could show its existence was a more or less inevitable
consequence of well-established physical laws and processes. Indeed, this is
essentially the claim that is made in the case of chaotic inflation. However
the problem is that the proposed underlying physics has not been tested, and
indeed may be untestable. There is no evidence that the postulated physics
is true in this universe, much less in some pre-existing metaspace that
might generate a multiverse. Thus belief in the validity of the claimed
physics that could lead to such consequences is just that, a belief - it is
based on unproved extrapolation of established physics to vastly beyond
where it has been tested. The issue is not just that the inflaton is not
identified and its potential untested by any observational means - it is
also that, for example, we are assuming quantum field theory remains valid
far beyond the domain where it has been tested, and we have faith in that
extreme extrapolation despite all the unsolved problems at the foundation of
quantum theory, the divergences of quantum field theory, and the failure of
that theory to provide a satisfactory resolution of the cosmological
constant problem.

\subsection{Observations and probabilities}

The `doomsday argument' has led to a substantial literature on relating
existence of universe models to evidence, based on analysis of
probabilities, often using a model of choosing a ball randomly from an urn,
and of associated selection effects (see e.g. Bostrom 2002). However usually
these models either in effect assume an ensemble exists, or else are content
to deal with potentially existing ensembles rather than actually existing
ones (see e.g. Olum 2002). That does not deal with the case at hand. One
would have to extend those arguments to trying to decide, on the basis of a
single ball drawn from the urn, as to whether there was one ball in the urn
or an infinite number. It is not clear to us that the statistical arguments
used in those papers leads to a useful conclusion in this singular case,
which is the case of interest for the argument in this paper.

In any case, in the end those papers all deal just with observational
probabilities, which are never conclusive. Indeed the whole reason for the
anthropic literature is precisely the fact that biophilic universes are
clearly highly improbable within the set of all possible universes (see e.g.
the use of Anthropic arguments as regards the value of $\Lambda $ referred
to in Section 5.2). We are working in a context where large improbabilities
are the order of the day. Indeed that is why multiverse concepts were
introduced in the first place - to try to introduce some form of scientific
explanation into a context where the probabilities of existence of specific
universe models preferred by observation are known to be very small.

\subsection{Observations and disproof}

Despite the gloomy prognosis given above, there are some specific cases
where the existence of a chaotic inflation (multi-domain) type scenario can
be disproved. These are when we live in a `small universe' where we have
already seen right round the universe (Ellis and Schreiber 1986,
Lachieze-Ray and Luminet 1995) for then the universe closes up on itself in
a single FLRW-like domain and so no further such domains that are causally
connected to us in a single connected spacetime can exist.

This `small universe' situation is observationally testable, and indeed it
has been suggested that the CBR\ power spectrum might already be giving us
evidence that this is indeed so, because of its lack of power on the largest
angular scales (Luminet et al, 2003). This proposal can be tested in the
future by searching for identical circles in the CMB sky. That would
disprove the usual chaotic inflationary scenario, but not a true multiverse
proposal, for that cannot be shown to be false by any observation. Neither
can it be shown to be true.

\section{Conclusion}

The introduction of the multiverse or ensemble idea is a fundamental change
in the nature of cosmology, because it aims to challenge one of the most
basic aspects of standard cosmology, namely the uniqueness of the universe
(see Ellis 1991, 1999 and references therein). However previous discussions
have not made clear what is required in order to define a multiverse,
although some specific physical calculations have been given based on
restricted low-dimensional multiverses. The aim of this paper is to make
clear what is needed in order to properly define a multiverse, and then
examine some of the consequences that flow from this.

Our fundamental starting point is the recognition that there is an important
distinction to be made between possible universes and realised universes,
and our main conclusion is that a really existing ensemble or multiverse is
not \textit{a priori} unique, nor uniquely defined. It must somehow be
selected for. We have pointed out a clear distinction between an ensemble of
possible universes $\mathcal{M}$,\thinspace and an ensemble of really
existing universes, which is envisioned as generated by the given primordial
process or action of an overarching cosmic principle. These effectively
select a really existing multiverse from the possibilities in $\mathcal{M}$,
and, as such, effectively define a distribution function over $\mathcal{M}$.
Thus, there is a definite causal connection, or \textquotedblleft law of laws
\textquotedblright , relating all the universes in these multiverses. It is
this really existing ensemble of universes, \textit{not} the ensemble of all
possible universes, which provides the basis for anthropic arguments.
Anthropic universes lie in a small subset of $\mathcal{M}$, whose
characteristics we understand to some extent. It is very likely that the
simultaneous realisation of \emph{all} the conditions for life will pick out
only a very small sector of the parameter space of all possibilities:
anthropic universes are fine-tuned.

The fine-tuning problem is very controversial. Two counter-attacks maintain
that there is no fine-tuning problem, so it is not necessary to construct
solutions to it by employing the multiverse idea. The first promotes the
view that whatever happens will always be unlikely (any hand of cards is as
unlikely as any other). Thus, since it is just an example of chance, there
is nothing special about a universe that admits life. The counter response
is that the existence of life is quite unlike anything else in the physical
world -- its coming into being is not just like choosing one out of numerous
essentially identical hands of cards. It is like being transformed into an
entirely different higher level game, and so does indeed require
explanation. The second counter-attack argues that inflation explains the
current state of the universe, making its apparently unlikely state
probable. However, this move is only partially successful, since very
anisotropic or inhomogeneous models may never inflate. The counter response
is that this does not matter: however small the chances are, if it works
just once then that is sufficient to give a model close enough to the
standard FLRW cosmological models to be friendly to life. But this does not
account for the rest of the coincidences enabling life, involving particle
masses and the values of the fundamental constants. Perhaps progress in
quantum cosmology will in the future lead to some unique theory of creation
and existence that will guide the discussion. At present, uniqueness eludes
us.

Among those universes in which the necessary cosmic conditions for life have
been fulfilled is the subset of almost-FLRW universes which are possible
models of our own observable universe, given the precision of the
observational data we have at present. It is, however, abundantly clear that 
\textquotedblleft really existing multiverses\textquotedblright\ which can
be defined as candidates for the one to which our universe belongs are 
\textit{not} unique, and neither their properties nor their existence is
directly testable. The only way in which arguments for the existence of one
particular kind of multiverse would be scientifically acceptable is if, for
instance, there would emerge evidence (either direct or indirect) for the
existence of specific inflaton potential which would generate one particular
kind of ensemble of expanding universe domains.

Despite these problems, the idea of a multiverse is probably here to stay
with us - it is an important concept that needs exploration and elucidation.
Does the idea that `all that can exist, exists' in the ensemble context
provide an explanation for the anthropic puzzles?\ Yes it does do so. The
issue of fine tuning is the statement that the biophilic set of universes is
a very small subset of the set of possible universes; but if all that can
exist exists then there are universe models occupying this biophilic
subspace. However there are the following problems: (i) the issue of
realised infinities discussed above, (ii) the problem of our inability to
describe such ensembles because we don't know what all the possibilities are,
so our solution is in terms of a category we cannot fully describe, and
(iii) the multiverse idea is not testable or provable in the usual
scientific sense; existence of the hypothesized ensemble remains a matter of
faith rather than proof. Furthermore in the end, \ it simply represents a
regress of causation. Ultimate questions remain: Why this multiverse with
these properties rather than others? What endows these with existence and
with this particular type of overall order? What are the ultimate boundaries
of possibility -- what makes something possible, even though it may never be
realised?\ In our view these questions - Issues 1 and 2 discussed in this
paper - cannot be answered scientifically because of the lack of any
possibility of verification of any proposed underlying theory. They will of
necessity have to be argued in philosophical terms.

The concept of a multiverse raises many fascinating issues that have not yet
been adequately explored. The discussion given here on how they can be
described will be useful in furthering this endeavour.

\section*{Acknowledgements}

\noindent We thank A. Linde, A. Lewis, A. Malcolm, and J.P. Uzan for helpful
comments and references related to this work and an anonymous referee of an
earlier version for comments. GFRE and UK acknowledge financial support from
the University of Cape Town and the NRF (South Africa).


\noindent

\begin{thebibliography}{99}
\bibitem{bos02}  Bostrom, N. 2002, \textit{Anthropic Bias} (Routledge).

\bibitem{bar}  Barrow, J\ D\ and Tipler, F\ J 1986, \textit{The Cosmic
Anthropological Principle }(Oxford University Press).

\bibitem{dav03}  Davies, P. 2003, \textquotedblleft Multiverse or Design?
Reflections on a `Third Way'\textquotedblright , unpublished manuscript and
talk given at the Centre for Theology and the Natural Sciences, Berkeley,
California, March 22, 2003.

\bibitem{dra}  Drake, F. and Shostak, S. \textit{Astrobiology:\ The Search
for Life in Space} (Cambridge University Press).

\bibitem{ehlrin89}  Ehlers, J. and Rindler, W. 1989, `A phase space
representation of Friedmann-Lema\^{\i}tre universes containing both dust and
radiation and the inevitability of a big bang'. \textit{Mon Not Roy Ast Soc} 
\textbf{238}, 503-521.

\bibitem{ell67}  Ellis, G F R 1967, ``The dynamics of pressure-free matter
in general relativity'', \textit{Journ Math Phys} 8, 1171-1194.

\bibitem{Ell71}  Ellis, G\ F\ R\ 1971, ``Relativistic Cosmology''. In 
\textit{\ \ \ \ General Relativity and Cosmology, Proc Int School of Physics
``Enrico Fermi'' } (Varenna), Course XLVII. Ed. R K Sachs (Academic Press),
104-179.

\bibitem{ell89}  Ellis, G F R 1991, ``Major Themes in the relation between
Philosophy and Cosmology''. \textit{Mem Ital Ast Soc} 62, 553-605.

\bibitem{ell99}  Ellis, G F R 1999, ``Before the Beginning: Emerging
Questions and Uncertainties''. In \textit{Toward a New Millenium in Galaxy
Morphology}, Ed. D Block, I Puerari, A Stockton and D Ferreira (Kluwer,
Dordrecht, 2000). \textit{Astrophysics and Space Science} 269-279: 693-720.

\bibitem{ellbru79}  Ellis, G. F. R. and Brundrit, G. B. 1979, 
\textquotedblleft Life in the infinite universe\textquotedblright . \textit{Qu 
Journ Roy Ast Soc} \textbf{20}, 37-41.

\bibitem{ellbruni}  Ellis, G F R and Bruni, M. 1989, \textquotedblleft A
covariant and gauge-free approach to density fluctuations in cosmology\textquotedblright .
\textit{Phys Rev} D40, 1804-1818.

\bibitem{ellmac}  Ellis, G.F.R. and MacCallum, M A H 1969, \textquotedblleft A
class of homogeneous cosmological models\textquotedblright . \textit{Comm
Math Phys} 12, 108-141.

\bibitem{ellma}  Ellis, G.F.R. and Maartens, R. 2003. ``The emergent
universe: Inflationary cosmology with no singularity''. gr-qc/0211082.

\bibitem{ellmu}  Ellis, G.F.R., Murugan, J. and Tsagas, C. G. 2003. ``The
Emergent Universe: An Explicit Construction''. gr-qc/0307112

\bibitem{ellsch}  Ellis, G.F.R, and Schrieber, G. 1986. \textit{Phys Lett}
A115, 97.

\bibitem{ellvan}  Ellis, G.F.R, and van Elst, H 1999, ``Cosmological
Models'' (Cargese Lectures 1998). In \textit{Theoretical and Observational
Cosmology}. Ed. M. Lachieze-Ray (Kluwer, Nato Series C: Mathematical and
Physical Sciences, Vol 541, 1999), 1-116. [gr-qc/9812046].

\bibitem{ellvan1}  van Elst, H. and Ellis, G.F.R. 1996, The covariant approach
to LRS perfect fluid spacetime geometries. \textit{Class. Quant. Grav.} 13:1099-1127

\bibitem{fish}  Fischer, A.E. and Marsden, J.E. 1979. ``The initial value
problem and the dynamical formulation of general relativity''. In \textit{\
General Relativity:\ An Einstein Centenary Survey}, Ed. S. W. Hawking and W.
Israel (Cambridge University Press).

\bibitem{garvil00}  Garriga J., Vilenkin A. 2000, `On likely values of the
cosmological constant', arXiv/astro-ph/9908115.

\bibitem{garvil01}  Garriga, J., and Vilenkin, A., 2001, `Many worlds in
one', \textit{Phys Rev} D64, 043511 (arXiv: gr-qc/0102010); gr-qc/0102090.

\bibitem{garvil02}  Garriga, J., and Vilenkin, A. 2002, arXiv:
astro-ph/0210358, arXiv:astro-ph/0210358

\bibitem{gutpi81}  Guth, A., and Pi, S. Y. 1981, \textit{Phys. Rev. Lett.}
49, 1110.

\bibitem{hew}  Hewitt, C. G., Horwood, \ J. T. and Wainwright, J. 2003,
``Asymptotic dynamics of the exceptional Bianchi cosmologies'',
\textit{Class. Quant. Grav.} \textbf{20} 1743-1756 [gr-qc/0211071].

\bibitem{hil64}  Hilbert, D. 1964, \textquotedblleft On the Infinite 
\textquotedblright, in \textit{Philosophy of Mathematics}, edited by Paul
Benacerraf and Hilary Putnam, Englewood Cliff, N. J. Prentice Hall, pp.
134-151.

\bibitem{hor}  Horwood, J. T., Hancock, M. J., The, D., and Wainwright, J.
2003. ``Late-time asymptotic dynamics of Bianchi VIII cosmologies'',
\textit{Class. Quant. Grav.} \textbf{20} \ 1757-1778 [gr-qc/0210031].

\bibitem{kallin02}  Kallosh R, Linde A.D. 2002, `M-theory, cosmological
constant and anthropic principle', arXiv/hep-th/0208157.

\bibitem{kirell03}  Kirchner, U., and Ellis, G. F. R. 2003, 
\textquotedblleft
A probability measure for FLRW\ models\textquotedblright\ \textit{Class.
Quantum Grav.} \textbf{20} 1199-1213.

\bibitem{lalu}  Lachieze-ray, M., and Luminet, J.P. 1995. \textit{Phys Rep} 
\textbf{254}, 135 [gr-qc/9605010].

\bibitem{lim}  Lim, W.C., van Elst, H., Uggla, C., and Wainwright, J. 2003,
"Asymptotic isotropization in inhomogeneous cosmology", gr-qc/0306118.

\bibitem{lin83}  Linde, A. D. 1983, \textit{Physics Letters} 129B, 177.

\bibitem{lin90}  Linde, A. D. 1990, \textit{Particle Physics and
Inflationary Cosmology}, Harwood Academic Publishers, Chur, Switzerland.

\bibitem{lin03}  Linde, A. D. 2003, \textquotedblleft Inflation, Quantum
Cosmology and the Anthropic Principle\textquotedblright , to appear in 
\textit{Science and Ultimate Reality: From Quantum to Cosmos}, honouring
John Wheeler's 90th birthday, J. D. Barrow, P. C. W. Davies and C. L.
Harper, editors, Cambridge University Press (2003) (arXiv:hep-th: 0211048
v2).

\bibitem{linetal94}  Linde, A. D., Linde, D. A., and Mezhlumian, A. 1994, 
\textit{Phys. Rev}. D49, 1783.

\bibitem{linmez93}  Linde, A. D., and Mezhlumian, A. 1993, 
\textquotedblleft
Stationary Universe\textquotedblright , arXiv:gr-qc/9304015.

\bibitem{lum}  Luminet, J.-P., Weeks, J.R., Riazuelo, A., Lehoucq, R., and
Uzan, J.-P. 2003, ``Dodecahedral space topology as an explanation for weak
wide-angle temperature correlations in the cosmic microwave background''. 
\textit{Nature}, xxx yyy.

\bibitem{olu02}  Olum, K. 2002, \textquotedblleft The doomsday argument and
the number of possible observers\textquotedblright , \textit{Phil Q} 52, 164
(gr-qc/0009081).

\bibitem{ree01}  Rees, M. J. 2001, \textit{Just Six Numbers: The Deep Forces
that Shape the Universe }(Basic Books). Rees,\ M.J. 2001 \textit{Our Cosmic
Habitat}, (Princeton University Press). M J\ Rees 2001, `Concluding
Perspective', astro-ph/0101268.

\bibitem{sci01}  Sciama, D. 1993, \textquotedblleft Is the universe unique?
\textquotedblright\ In \textit{Die Kosmologie der Gegenwart}, Ed G Borner
and J Ehlers (Serie Piper).

\bibitem{smo99}  Smolin, L. 1999, \textit{The Life of the Universe }(Oxford
University Press)\textit{.}

\bibitem{spi00}  Spitzer, R. J. 2000, \textquotedblleft Definitions of Real
Time and Ultimate Reality\textquotedblright , \textit{Ultimate Reality and
Meaning} \textbf{23} (No. 3), 260-276.

\bibitem{sta86}  Starobinsky, A. A. 1986, \textquotedblleft Stochastic De
Sitter (Inflationary) Stage in the Early Universe\textquotedblright , in 
\textit{Current Topics in Field Theory, Quantum Gravity and Strings}, eds.
H. J. de Vega and N. Sanchez (Springer, Heidelberg), p. 107.

\bibitem{sto03}  Stoeger, W. R. 2003, \textquotedblleft What is `the
Universe' that Cosmology Studies\textquotedblright , to be published in the
Ian Barbour Festschrift.

\bibitem{teg98}  Tegmark, M. 1998,\ \textquotedblleft Is the Theory of
Everything Merely the Ultimate Ensemble Theory?\textquotedblright\ \textit{\
Annal Phys} 270, 1-51 (arXiv.org/gr-qc/9704009).

\bibitem{teg03}  Tegmark, M. 2003, \textquotedblleft Parallel Universes
\textquotedblright , astro-ph/0302131 and \textit{\ Scientific American }
(May 2003), 41-51.

\bibitem{uggetal03}  Uggla, C., van Elst, H., Wainwright, J. and Ellis, G.
F. R. 2003, `The past attractor in inhomogeneous cosmology'. gr-qc/0304002.

\bibitem{vil95}  Vilenkin, A. 1995, \textit{Phys. Rev. Lett}. 74, 846.

\bibitem{wai}  Wainwright,\ J. and Ellis, G. F. R. (Eds) (1996). \textit{The
dynamical systems approach to cosmology}. Cambridge University Press.

\bibitem{wei97}  Weinberg, S. 1997, in \textit{Critical Dialogs in Cosmology}
, edited by Neil Turok, World Scientific, Singapore.

\bibitem{wei00}  Weinberg, S. 2000, The Cosmological Constant Problems',
astro-ph/0005265. `A Priori Probability Distribution of the Cosmological
Constant', \textit{Phys. Rev. }D61, 103505 (2000) (astro-ph/0002387).
\end{thebibliography}
\end{document}